\documentclass[11pt,a4paper]{article}
\usepackage{amsfonts}
\usepackage{amsmath}

\DeclareMathOperator*{\argmin}{arg\,min}
\usepackage{eurosym}
\usepackage[a4paper]{geometry}
\usepackage[table]{xcolor}
\usepackage{float}
\usepackage[caption = false]{subfig}
\usepackage{graphicx}
\usepackage{tikz}
\usepackage{bm}
\usepackage{natbib}
\usepackage[flushleft]{threeparttable}
\usepackage[pdfborder={000}]{hyperref}

\begin{document}

\title{Probabilistic Forecasting in Day-Ahead Electricity Markets: Simulating Peak and Off-Peak Prices}
\author{Peru Muniain\thanks{%
University of the Basque Country, UPV/EHU. E-mail: peru.muniain@ehu.eus.}
Florian Ziel\thanks{%
University of Duisburg-Essen,uni-DUE. E-mail: florian.ziel@uni-due.de.},
}
\date{}
\maketitle

\begin{abstract}
In this paper we include dependency structures for electricity price forecasting and forecasting evaluation. We work with off-peak and peak time series from the German-Austrian day-ahead price, hence we analyze bivariate data. We first estimate the mean of the two time series, and then in a second step we estimate the residuals. The mean equation is estimated by OLS and elastic net and the residuals are estimated by maximum likelihood. Our contribution is to include a bivariate jump component on a mean reverting jump diffusion model in the residuals. The models' forecasts are evaluated using four different criteria, including the energy score to measure whether the correlation structure between the time series is properly included or not. In the results it is observed that the models with bivariate jumps provide better results with the energy score, which means that it is important to consider this structure in order to properly forecast correlated time series.

\end{abstract}

\section{Introduction}

In the last few decades since the deregulation of electricity markets it has become increasingly important to capture uncommon features of electricity prices such as nonstorability, which makes electricity prices really volatile ( see \cite{weron2014electricity}). In this paper we use different time series models to forecast electricity by simulation and then evaluate those forecasts using various criteria with different properties. We believe it is crucial to take into account the dependency structures in order to properly forecast multivariate time series. 
The innovation in this paper is that we include dependency structures in some of the multivariate forecasting models and in one of the forecast evaluation criteria to show that the incorporation of the dependency structures substantially improves electricity price forecasts. The electricity prices we model and forecast in this paper are the peak and off-peak price series based on the German-Austrian day-ahead price. These time series are important for derivatives trading.

As mentioned above, electricity prices show special characteristics which are usually classified in the relevant literature ( see \cite{weron2014electricity} and \cite{ziel2016forecasting}). Specifically, these properties are i) mean reverting behavior; ii) seasonal behavior; iii) time dependent volatility; iv) price spikes; and v) cross-period effects (e.g. night hours influence day-time hours even though they take place on the same day).
All these aspects are known in the literature, but there is no electricity price forecasting model which incorporates all of them. 
For instance, \cite{karakatsani2008forecasting} cover all the above effects except interaction effects, \cite{ziel2015efficient} consider all effects except price spike effects.
We propose electricity price models which incorporate all the said effects into a probabilistic electricity price forecasting framework.

A two-step approach is used to forecast prices. In the first step the conditional mean model is estimated: the mean must be properly estimated so that the residuals have a zero mean. Therefore, in the mean equation all the seasonal properties must be included. Accordingly, \cite{uniejewski2016automated} and \cite{ziel2018day} propose mean equations with autoregressive, non-linear effects and seasonal effects. Once the conditional mean model is properly estimated we proceed to estimate the residuals, which must have a zero mean so the models differ in the structures of the standard deviation. We consider mean reverting jump diffusion models (MRJD) such as the model included by \cite{seifert2007modelling} and \cite{ioannou2018effect} applied to electricity prices. The MRJD model is an Ornstein–Uhlenbeck (OU) process proposed by \cite{uhlenbeck1930theory}. Unlike \cite{weron2008market} and \cite{cartea2005pricing}, where the jump component is first estimated and then an OU process is assumed in the continuous part; we first estimate the mean model and then we assume a MRJD structure in the residuals. \cite{weron2014electricity} offers a good review of MRJD models applied to electricity price forecasting. As mentioned above, our interest is the dependency structure between different time series, which we include by assuming correlated jump occurrence processes, a procedure we believe has never been used before. To obtain a correlated jump we focus on the bivariate Bernoulli process, proposed by \cite{dai2013multivariate}.

Once the models are estimated electricity prices are simulated and forecast, then those forecasts and their paths are evaluated using different criteria. In this article we use four different criteria: mean absolute error (MAE), mean square error (MSE), pinball score (PB, also known as quantile loss) and energy score (ES). The first two are the most widely used in the literature of forecasting evaluation; for instance, \cite{keles2012comparison} apply MSE to evaluate the electricity price forecasts from a model including spikes and other structures such as ARIMA and GARCH. \cite{voronin2014hybrid} use the MSE and MAE criteria to evaluate the performance of different electricity price forecasts in the NORDPOOL market. In this paper we focus more on the PB and ES as we are interested in the performance capturing the whole distribution and how the different models capture dependency structures. The PB has been applied by \cite{maciejowska2016hybrid}, \cite{dudek2016multilayer}, and \cite{juban2016multiple}, all involving an electricity price forecasting competition with the PB used to check performance, as the objective was to approximate the forecast distribution. The ES has not been applied to electricity price forecasts so far, \cite{weron2019electricity}. However, it has been applied a few times in the energy forecasting context, e.g. in \cite{pinson2012evaluating} for wind power forecasting. The ES is built up as per \cite{gneiting2007strictly} and then applied to our time series. We pay more attention to this score because it takes into account dependency structures. As mentioned above, our contribution is to include correlation structures in the models as well as in the evaluation. 
Then, to check whether the differences between the forecasting performances of the models in pairs are significant or not, the Diebold-Mariano (DM) test is applied. 

The rest of the paper is organized as follows: Section \ref{sec:data} explains the data and highlights the relationships to the derivative markets, Section \ref{sec:model} introduces the models, Section \ref{sec:estimation} explains the estimation methods and how the forecasts are generated, Section \ref{sec:evaluation} describes the evaluation criteria, Section \ref{sec:results} discusses the results, and Section \ref{sec:conclusion} summarizes our results, and outlines the most important facts.

\section{Data} \label{sec:data}
\subsection{Motivation}

As mentioned in the introduction, we focus on off-peak and peak price series from the EPEX market because they are relevant for derivative trading, 
especially future products.
On the European Energy Exchange (EEX) different future products for electricity with cash settlement for the German/Austrian delivery zone are traded.
They are base, off-peak and peak price products (also known as Phelix) traded at EEX. The underlying of these products are based on the hourly German/Austrian EPEX day-ahead  electricity prices. The Phelix base product is simply calculated as the mean of all hourly EPEX prices in the delivery period. For example, the underlying of Phelix base week future contracts are calculated as the mean of the 168 hourly prices from Monday 0:00-1:00 to Sunday 23:00-24:00.
For Phelix peak products the underlying is the mean of the day-ahead price from the $9^{\text{th}}$ hour of the day to the  $20^{\text{th}}$ (12 hours in total) on Monday to Friday. Thus, for Phelix peak week futures contracts the underlying are computed as a mean of the $5\times12=60$ hourly mean prices for the peak hours from Monday to Friday. The remaining $168-60=108$ hours would be the underlying for Phelix off-peak week future products.
However, Phelix off-peak products are only available for longer delivery periods (month, quarter and annual) and are rather illiquid. Therefore,
the primarily focus for traders is on the Phelix base and peak products.

As traders focus on base and peak products it makes sense to concentrate on forecasting the corresponding underlyings.
However, the fact that definition of the Phelix peak products depends on the day of the week makes the modeling a bit cumbersome.
Intuitively, it makes sense to model and forecast the daily base price (the mean of the 24 hourly prices) and the daily peak price (the mean of the 12 prices 8:00-9:00 to 19:00-20:00). Of course, for trading Phelix peak products a forecast for Saturday and Sunday peak prices is not relevant. Nonetheless, it is more convenient to model the peak price in the above mentioned manner to preserve the time series structure.
However, the base and peak time series are partially based on the same prices, in fact the peak prices. But, from the modeling perspective 
it is more convenient to have less correlated data. 
This linear dependency can be reduced easily by modeling the daily peak and off-peak prices as they are computed based on completely different hourly prices. If we are interested in a base price, we may obtain it directly by averaging the daily off-peak and peak prices.
Hence, it is completely sufficient to model the base and peak prices for trading purposes. Thus, we proceed to analyze the  above mentioned time series henceforth.

%

%

Finally, we would like to mention that it would be more informative to have a model for the 24 hourly electricity prices than just a model for the peak and 
off-peak prices. 
The problem with these models in the considered probabilistic forecasting setup are the computational burden,
as there would be too many variables to estimate and we would not be able to optimize the models. However, these forecasts would not add any information regarding derivatives markets because hourly prices are not traded in these markets.

\subsection{Description}

The considered electricity price data starts on 1 st January 2014 and ends on 31 st December 2017. It is measured in EUR/MWh. To calculate the higher moments and the dependencies, use the following notation;

$$
\text{m}_{i,j} = \mathbb{E}\left[\left(\frac{Y_{d,1}-\mu_{Y_{d,1}}}{\sigma_{Y_{d,1}}}\right)^i \left(\frac{Y_{d,2}-\mu_{Y_{d,2}}}{\sigma_{Y_{d,2}}}\right)^j\right],
$$
\noindent where $Y_{d,1}$ and $Y_{d,2}$ refer to off-peak and peak time series with their means $\mu_{Y_{d,1}}$ and $\mu_{Y_{d,2}}$ and standard deviations $\sigma_{Y_{d,1}}$ and $\sigma_{Y_{d,2}}$. We show below the sample statistics (of the input data), but these may not be good estimators for the corresponding statistical counter-parts. However, under some mixing assumptions (e.g. weakly periodically stationary) the sample mean/variance/skewness/kurtosis/etc converge to the corresponding counterpart. Additionally, we would like to point out the fact that if the time series are bounded, and this is our case, then all moments exist. The sample descriptive statistics for both time series are shown below:

\begin{table}[ht]
\caption{Descriptive statistics of off-peak and peak prices}
\label{tab:descriptive}
\resizebox{\textwidth}{!}{
\centering
\begin{tabular}{|rrrrrrrrr|}
  \hline
              & mean  & sd    & median& min    & max & cor & skew & coskew \\ 
  \hline
off-peak   & 28.30 & 8.74  & 29.36 & -56.38 & 73.66  & 0.80 & -1.61 & -0.59 \\  
  peak     & 35.48 & 13.79 & 35.09 & -45.27 & 130.18 & 0.80 & 0.48 & 0.01 \\  
   \hline
\end{tabular}
}
\begin{tablenotes}
\footnotesize
\item The number of days is 1461, and sd is standard deviation. For the higher moments and dependencies, cor refers to correlation, which is an estimator of m$_{1,1}$ (same for both time series), skew to skewness with our notation estimated value of m$_{3,0}$ and m$_{0,3}$ for off-peak and peak, respectively. Similarly, coskew references to coskewness which gives estimations of m$_{2,1}$ and m$_{1,2}$. 
\end{tablenotes}  
\end{table}

As expected, Table \ref{tab:descriptive} shows that the mean and the standard deviation are higher in the peak time series. As the volatility is higher the range for the peak series is higher than that of the off-peak time series. The correlation shows quite a high positive linear relationship between the two time series. The skewness shows that the off-peak series is clearly asymmetric and that the peak series is slightly asymmetric. The coskewness coefficients show how the variance of one time series and the mean of the other are related. As observed in Table \ref{tab:descriptive}, the relationship between the off-peak central variance and the peak central mean is stronger than the other way round; in the case of m$_{2,1} = -0.59$, this means that the higher the value of the peak series the lower the variance of the off-peak series. In view of these results it can be concluded that none of the time series follows a normal distribution pattern.

\begin{figure}[h!]
\subfloat{\includegraphics{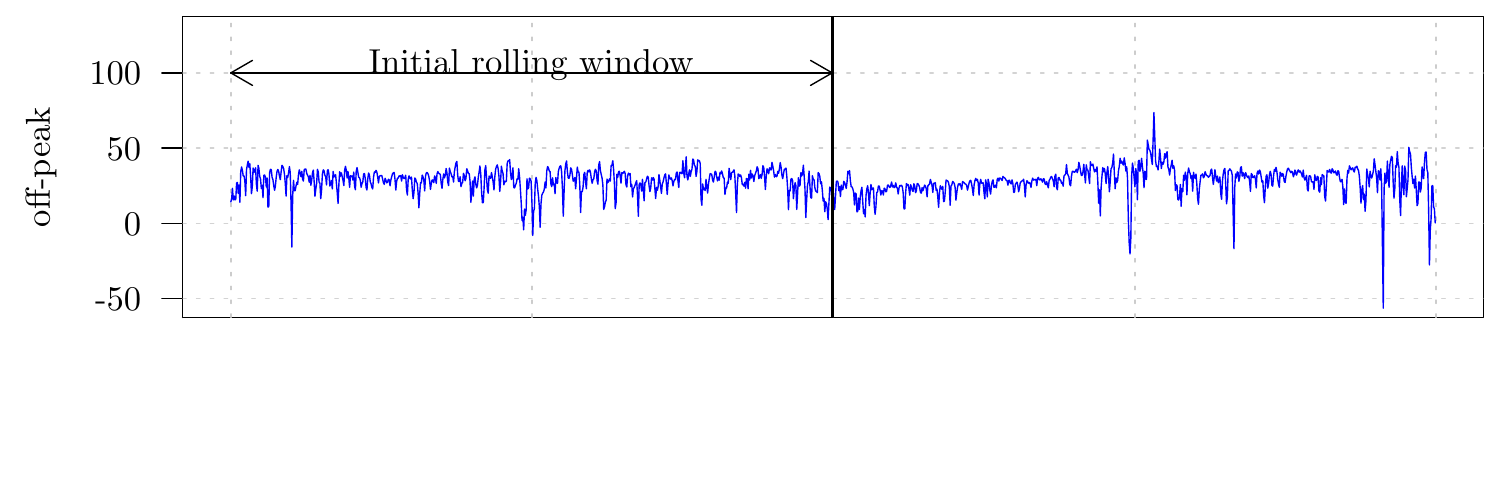}}\\[-13ex] 
\subfloat{\includegraphics{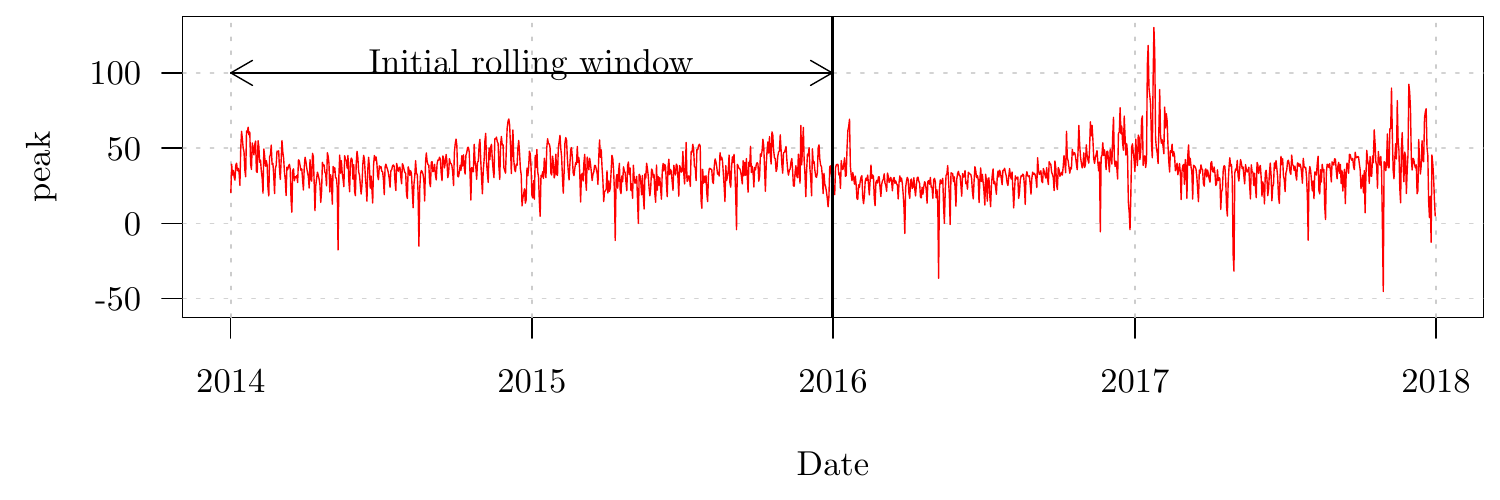}}
\caption{off-peak and peak time series}
\label{fig:offpeakpeak}
\end{figure}

Figure \ref{fig:offpeakpeak} shows the off-peak and peak time series through our sample. The first two years are used only for estimation purposes and the last two years are first predicted and then used as observations of the following rolling windows. How the rolling windows are developed is explained in Section \ref{sec:estimation}. Figure \ref{fig:offpeakpeak} is divided in two to emphasize this aspect. It is observed in Figure \ref{fig:offpeakpeak} that the volatility was higher  at the beginning of 2017  and also at the end of the year. As can be observed in Table \ref{tab:descriptive} and in Figure \ref{fig:offpeakpeak}, the volatility is higher, and so is the mean in the peak series compared to the off-peak figures. However, generally the trend in the graphs is quite similar, as shown by the correlation coefficient. In both cases there is evidence of volatility clustering and spikes.

\begin{figure}[h!]
 \subfloat[off-peak histogram and density]{\includegraphics{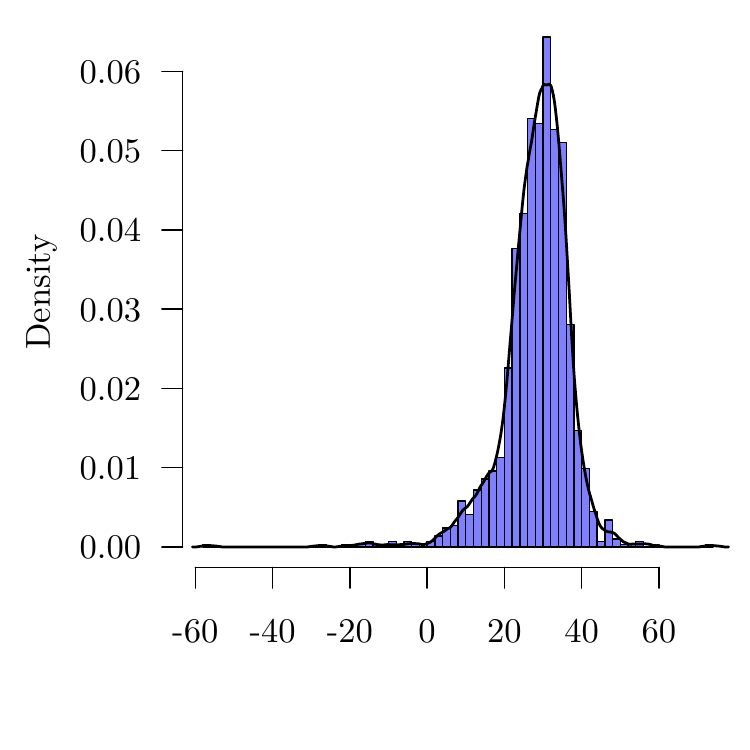}\label{fig:histoffpeak}}\hfill
 \subfloat[peak histogram and density]{\includegraphics{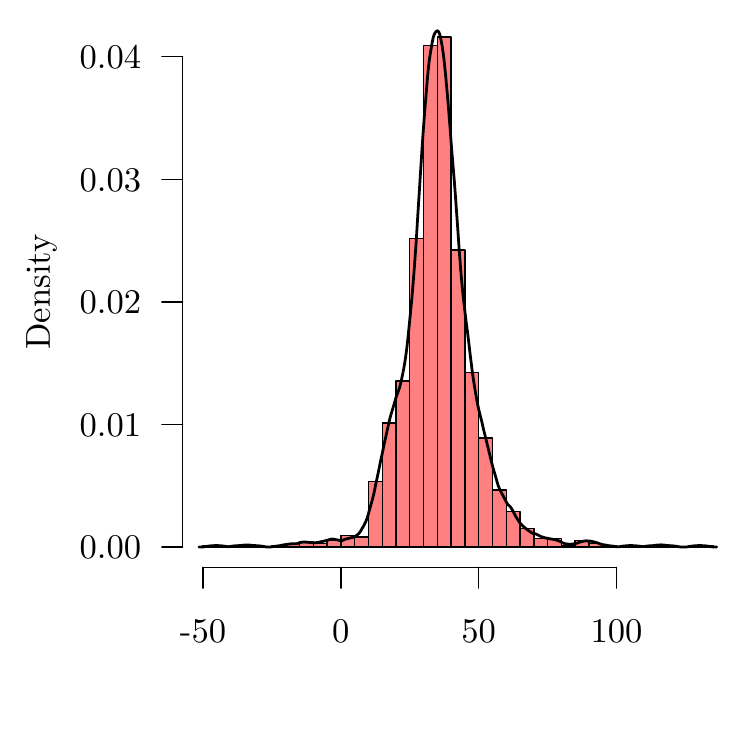}\label{fig:histpeak}}\\[-1ex] 
 \subfloat[Scatter  plot]{\includegraphics{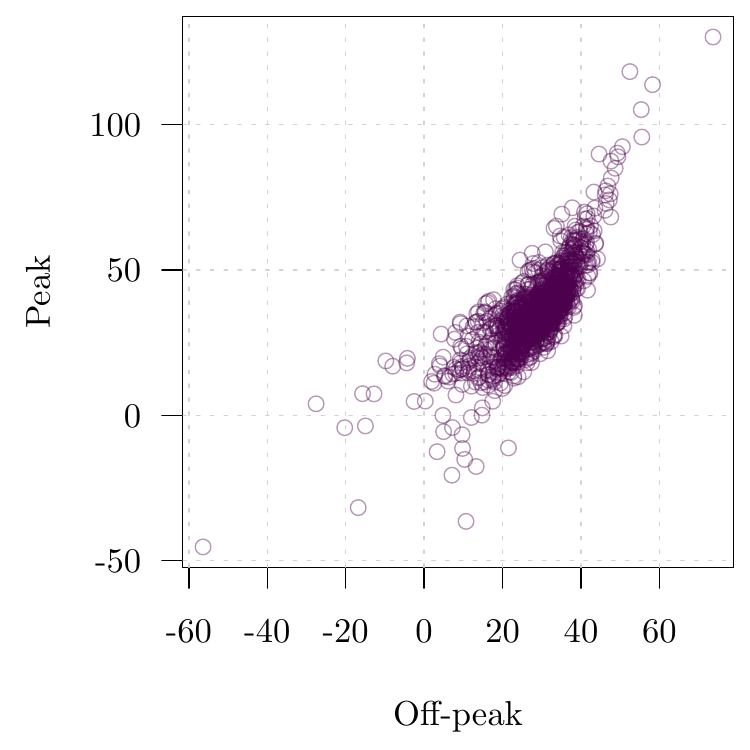}}
 \subfloat[Bivariate density]{\includegraphics{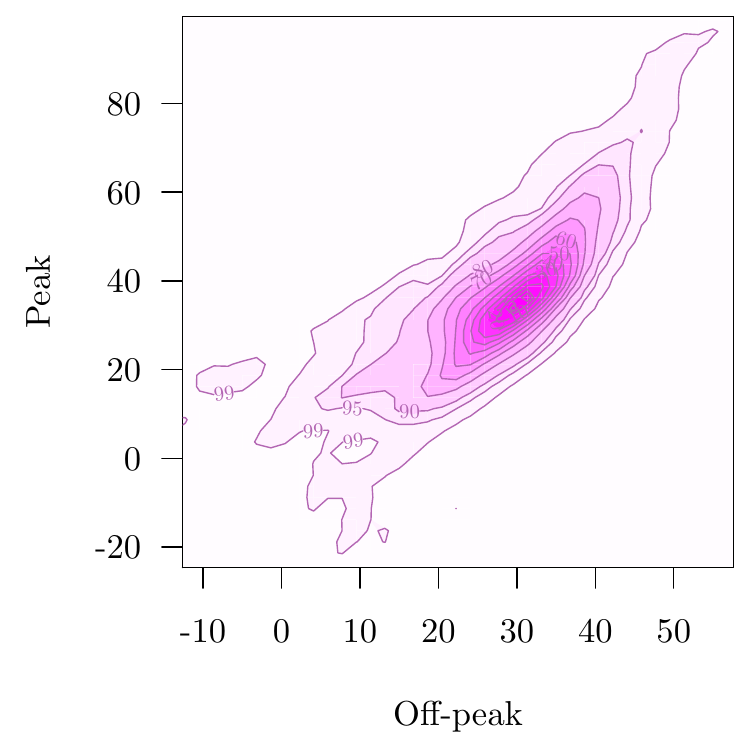} \label{fig:bidens}}
\caption{Histograms and densities of off-peak and peak time series}
\label{fig:hist}
\end{figure}

The histograms and density functions in Figure \ref{fig:hist} show the distribution of the two time series. It may be observed that both series have heavy tails and the asymmetry is more pronounced in the off-peak series. In both cases there is evidence of spikes, which are rare events where the price is extremely low or high. Regarding the scatter plot, a strong and complex correlation between the two is confirmed, which leads us to include correlation structures in our models. In the models we propose, the correlation is not included only in the continuous part of the variation but also in the jump occurrence process, as the depicted graphs show. From the scatter plot it is also possible to observe the bivariate density, which shows how the scatter plot is distributed. In Figure \ref{fig:bidens}, the darker colors show the higher quantiles of the distribution. The bivariate distribution confirms the intuition of the scatter plot, where the darker areas are those where there are more points.  In both graphs - the scatter plot and the bivariate distribution - one may observe that the spread is higher in lower values than in higher values.. The correlation coefficient for the values when the off-peak price is lower than 30€ is 0.65, while when the off-peak price is higher than 30 the correlation coefficient is 0.84. This is an example of the complicated dependency structure.

\section{Models} \label{sec:model}

The models that we analyze in this paper are two step models. In the first step we estimate the mean equation and in the second we study the residuals from the previous step.

For the sake of simplification we define $\bm{Y}_d = (Y_{d,1}, Y_{d,2})'$ as the bivariate vector of the off-peak and peak prices, so index $1$ corresponds to the off-peak price and index $2$ to the peak price.

\subsection{ARX type models}

In this subsection we introduce the conditional mean model that we assume. The mean equation is based on the mean models proposed in \cite{uniejewski2016automated} and \cite{ziel2018day}. To calculate the mean equation we assume a model with autoregressive structure with exogenous variables (ARX) for the peak and off-peak series. The ARX model was shown to perform really well in forecasting electricity prices in \cite{uniejewski2016automated} and \cite{ziel2018day}. We consider the mean model for the two time series as:

\begin{align}
Y_{d,i} =& \beta_{i,0} + \sum_{k=1}^8 \left(\beta_{i,k,1}Y_{d-k,1}+ \beta_{i,k,2}Y_{d-k,2}\right)  \nonumber \\
&+ \sum_{k=1}^7 \left[\left(\beta_{i, k,3}  + \beta_{i,k,4} Y_{d-1,1} + \beta_{i,k,5} Y_{d-1,2} \right)\text{DoW}_d^k\right]+\epsilon_{d,i} \label{eq:mean}
\end{align}

\noindent where $i=1,2$ and $\text{DoW}_d^j$ is a day of the week dummy of day $j$ at day $d$ such that e.g. $\text{DoW}_d^1$ is $1$ if $d$ falls on a Monday, 
$\text{DoW}_d^2 = 1$ if $d$ is on Tuesday  etc. The residuals are $\epsilon_{d,1}$ and $\epsilon_{d,2}$, and by construction the mean of the two terms must be 0. 
The model has in total $p=1 + 2\times 8 + 3\times7 = 38 $ parameters with corresponding parameter vector $ \bm{\beta}$.
Obviously, model \eqref{eq:mean} is a linear model that can be written as
\begin{align}
Y_{d,i} = \bm{X}_{d,i}' \bm{\beta}_i + \epsilon_{d,i}
\label{eq:mean_lm}
\end{align}
where $\bm{X}_{d,i} $ and $ \bm{\beta}_i$ are $p$-dimensional.

The error terms are considered to be distributed as:
\begin{align}
\bm{\epsilon}_d \sim \mathcal{N}_2(\bm{0},\,\bm{\Sigma})
\label{eq_res_binorm} 
\end{align}
\noindent where $\bm{\epsilon}_d=(\epsilon_{d,1},\epsilon_{d,2})'$, $\bm{0} = (0,0)'$ and $\bm{\Sigma}$ is the covariance matrix of $\bm{\epsilon}_d$.

Model \eqref{eq:mean} covers the major characteristics of electricity prices, especially mean reverting properties, seasonal structure, and cross-period effects.
Only volatility and price spikes are not captured by the structure assumed.
Hence, for all the remaining models we consider the same mean equation, but modify the error model \eqref{eq_res_binorm} to capture the missing effects. 


\subsection{ARX type models with independent jumps in the residuals}

In this subsection we explain the ARX-IJ model. We consider MRJD in each residual independently. This is the standard OU process applied in electricity price forecasting, and has been applied several times, e.g. in \cite{keles2012comparison} and widely discussed in \cite{weron2014electricity}.  Jump diffusion models are accurate for capturing price spikes as observed in the tails of Figures \ref{fig:histoffpeak} and \ref{fig:histpeak}.

After Euler discretization the model is written as follows:
\begin{align*}
\bm{\epsilon}_d =& \bm{\epsilon}_{d,cont} + \bm{B}_d  \bm{\epsilon}_{d,jump}\\
\bm{\epsilon}_{d,cont} \sim& \mathcal{N}_2(      - \bm{\Lambda} \bm{\mu},\,\bm{\Sigma}) \\ 
\bm{\epsilon}_{d,jump} \sim& \mathcal{N}_2( \bm{\mu} ,\,\bm{\Gamma})\\ 
&\text{where} \ \ \bm{\mu} =
\left( {\begin{array}{c}
   \mu_1 \\
   \mu_2 \\
  \end{array} } \right) ,
  \bm{\Sigma}=\left( {\begin{array}{cc}
   \sigma_{1}^2 & \rho\sigma_{1}\sigma_{2}  \\
   \rho\sigma_{1}\sigma_{2} & \sigma_{2}^2 \\
  \end{array} } \right) , \bm{\Lambda} = 
  \left( {\begin{array}{cc}
   \lambda_1 & 0  \\
   0 & \lambda_2 \\
  \end{array} } \right) 
  \\
 &\bm{B}_d = \left( {\begin{array}{cc}
   b_{d,1} & 0\\
    0 & b_{d,2} \\
  \end{array} } \right)  \ \text{with} \ b_{d,i} \sim \text{Ber}(\lambda_{i}) \ \text{for} \ i = 1,2 , 
  \text{and} \ \ \bm{\Gamma}=\left( {\begin{array}{cc}
   \gamma_{1}^2 & 0 \\
   0 & \gamma_{2}^2 \\
  \end{array} } \right) ,
\end{align*}
\noindent where $\bm{\epsilon}_{d,cont}$ is the continuous part of the error term and $\bm{\epsilon}_{d,jump}$ is the jump component. 
$\lambda_i$ is the probability of jumps, $\text{Ber}$ is the Bernoulli distribution, $\mu_i$ is the mean size of the jump, $\gamma_i$ is the standard deviation of the jump and $\sigma_i$ is the standard deviation of the continuous part all of them defined for $i=1,2$. We need the terms $-\lambda_{1} \mu_{1}$ and $-\lambda_2 \mu_2$ in the continuous term in order to ensure that the mean of $\epsilon_{d,1}$ and $\epsilon_{d,2}$ is 0, as it must be by construction. In this model we assume that the Bernoulli random variables $b_{d,1}$ and $b_{d,2}$ are independent. The conditional error term $\bm{\epsilon}_d | \bm{B}_d = \bm{\epsilon}_{d,cont} + \bm{B}_d \bm{\epsilon}_{d,jump}|\bm{B}_d$ is distributed as follows:
\begin{equation}
\bm{\epsilon}_d  | \bm{B}_d \sim \mathcal{N}_2 ( \bm{0},\,\bm{\Sigma}+\bm{B}_d\bm{\Gamma} \bm{B}_d').
\label{eq:dist}
\end{equation}
For the unconditional distribution of $\bm{\epsilon}_d$ 
first note that with
$\text{Var}[XY] = \mathbb{E}[X]^2\text{Var}[Y] + \text{Var}[X] \mathbb{E}[Y]^2 + \text{Var}[X]\text{Var}[Y]$
the following holds:
\begin{align}
\text{Var}[b_i \epsilon_{d,jump,i}] 
&=
 \mathbb{E}[b_i]^2\text{Var}[\epsilon_{d,jump,i}] + \text{Var}[b_i] \mathbb{E}[\epsilon_{d,jump,i}]^2 + \text{Var}[b_i]\text{Var}[\epsilon_{d,jump,i}] \nonumber \\
&=
 \lambda_i^2 \gamma_i^2 + \lambda_i(1-\lambda_i)  \mu_i^2 + \lambda_i(1-\lambda_i) \gamma_i^2 
 =  \lambda_i(  (1-\lambda_i)\mu_i^2 + \gamma_i^2) .
 \end{align}
Thus, as a result of the independence of all occurring random variables, it holds that
\begin{align}
 \text{Var}[ \bm{\epsilon}_d ]
 = \bm{\Sigma} + \text{Var}[\bm{B}_d  \bm{\epsilon}_{d,jump} ] 
= \bm{\Sigma} + \bm{\Lambda}  ( (\bm{I} -\bm{\Lambda} ) \text{Diag}(\bm{\mu})^2 + \text{Diag}(\bm{\Gamma}) )
\end{align}

as $\text{Cov}[b_{d,1} \epsilon_{d,jump,1}, b_{d,2} \epsilon_{d,jump,2} ] =0$.\\

However, it is clear that $\bm{\epsilon}_d$ does not follow a bivariate normal distribution pattern.

\subsection{ARX type models with bivariate jumps in the residuals}

The next model that we introduce, the ARX-BiJ model, is related to the previous one as it is based on an MRJD structure, but in this case the jump component is assumed to be bivariate ( more precisely bivariate Bernoulli). This dependency structure in the jump is one of our contributions to the literature.
We further assume that the jump sizes can be correlated.
We write the model as
\begin{align*}
\bm{\epsilon}_d =& \bm{\epsilon}_{d,cont} + \bm{B}_d  \bm{\epsilon}_{d,jump} \ \ \text{ with } \ \
\bm{\epsilon}_{d,cont} \sim \mathcal{N}_2(      - \bm{\Lambda} \bm{\mu},\,\bm{\Sigma})  \ \ \text{ and } \ \ 
\bm{\epsilon}_{d,jump} \sim \mathcal{N}_2( \bm{\mu} ,\,\bm{\Gamma}),\\ 
&\text{where } \bm{\mu} =
\left( {\begin{array}{c}
   \mu_1 \\
   \mu_2 \\
  \end{array} } \right),
  \ \ \bm{\Sigma}=\left( {\begin{array}{cc}
   \sigma_{1}^2 & \rho\sigma_{1}\sigma_{2} \\
   \rho\sigma_{1}\sigma_{2} & \sigma_{2}^2 \\
  \end{array} } \right),
  \bm{\Gamma}=\left( {\begin{array}{cc}
   \gamma_{1}^2 & \varrho\gamma_{1}\gamma_{2} \\
   \varrho\gamma_{1}\gamma_{2} & \gamma_{2}^2 \\
  \end{array} } \right)
  \\
  &\bm{B}_d = \left( {\begin{array}{cc}
   b_{d,1} & 0 \\
   0 & b_{d,2} \\
  \end{array} } \right) \ \text{ with } \text{diag}(\bm{B}_d) \sim \text{\textbf{Ber}}_2(\bm{P}) \\ & \text{with probabilities }  \bm{P} = \left( {\begin{array}{cc}
   p_{0,0} & p_{1,0} \\
   p_{0,1} & p_{1,1} \\
  \end{array} } \right), 
   \bm{\Lambda} = \left( {\begin{array}{cc}
   \lambda_{1} &0\\
  0& \lambda_{2} \\
  \end{array} } \right) = \left( {\begin{array}{cc}
   p_{1,0}+p_{1,1} \\
   p_{0,1}+p_{1,1} \\
  \end{array} } \right)
\end{align*}

As mentioned above, we now assume that the arrivals of the jumps are bivariate $\text{\textbf{Ber}}_2(\bm{P})$ distributed with probabilities $\bm{P}$.
In this case, $p_{0,0}$ is the probability of no jump, $p_{1,0}$ is the probability of a jump occurring only in the off-peak component, $p_{1,1}$ is the probability of there being a jump in both components at the same time, and $p_{0,1}$ is the probability of a jump occurring only in the peak component. Therefore, $ \lambda_2 = p_{0,1}+p_{1,1}$ is the total probability of jumps in the peak component, and $\lambda_1= p_{1,0}+p_{1,1}$ is the equivalent probability in the off-peak series. The condition $p_{1,0}+p_{1,1}+p_{0,1} + p_{0,0}=1$ must hold. Unlike the previous model, in this case $b_{d,1}$ and $b_{d,2}$ are not independent: they must coincide in no jump with probability $p_{0,0}$ and in jump with probability $p_{1,1}$. For the bivariate Bernoulli setting we follow \cite{dai2013multivariate}. Assuming a bivariate jump process. we capture dependency structure in the continuous component as well as in the jump component.

\subsection{ARX type models with bivariate jumps in the residuals with no constant mean.}

This subsection presents the ARX-BiJ-$\mu_d$ model. This model is very similar to the previous one, but in this case the mean of the jump is assumed to depend on the price observed previously. In order not to make things too tedious for the reader we only note those points that differ from the previous model, i.e.:
\begin{align}
&\bm{\mu}_{d} = \bm{\mu}_{0} + \bm{\mu}_{1}\bm{Y}_{d-1} \text{ and } \bm{\epsilon}_d = \bm{\epsilon}_{d,cont} + \bm{B}_d  \bm{\epsilon}_{d,jump} \label{eq:bimean}\\ 
& \text{with } 
\bm{\epsilon}_{d,cont} \sim \mathcal{N}_2(      - \bm{\Lambda} \bm{\mu}_d,\,\bm{\Sigma}), 
\text{diag}(\bm{B}_d) \sim \text{\textbf{Ber}}_2(\bm{P})  \text{ and }  
\bm{\epsilon}_{d,jump} \sim \mathcal{N}_2( \bm{\mu}_d ,\,\bm{\Gamma}),\nonumber
\end{align}

In this model we seek to capture the effect of the previous price on the mean of the jump component. 

\subsection{ARX type models with CCC-GARCH}

The next model (ARX-GARCH) considers bivariate constant conditional correlation GARCH (CCC-GARCH) structures, as first introduced by \cite{bollerslev1990modelling}. We follow \cite{silvennoinen2009multivariate} in their implementation:
\begin{align*}
\epsilon_{d,i}=&\sigma_{d,i} z_{d,i}\\
\sigma_{d,i}^2 =& \alpha_{0,i} + \alpha_{1,i}\epsilon_{d-1,i}^2+\alpha_{2,i}\sigma_{d-1,i}^2 \ \text{for} \ i=1,2,\\
\bm{\epsilon}_d \sim& \mathcal{N}_2(\bm{0},\,\bm{\Sigma}_d)\ \ \text{where} \ \ \bm{\Sigma}_d=\left( {\begin{array}{cc}
   \sigma_{d,1}^2 & \rho\sigma_{d,1}\sigma_{d,2} \\
   \rho\sigma_{d,1}\sigma_{d,2} & \sigma_{d,2}^2 \\
  \end{array} } \right)
\end{align*}
\noindent where $z_{d,1}$ and  $z_{d,2}$ are independent white noises with a standard deviation of 1. The parameters of the GARCH structure must fulfill $\alpha_{0,i},\alpha_{1,i},\alpha_{2,i}>0$ and $\alpha_{1,i}+\alpha_{2,i}<1$ conditions in order for the time series to be stationary. 
In this model we assume that the correlation between the two time series is constant and there are no cross-dependencies between the volatility series. Structures of this type are often used in the literature to forecast multivariate time series, for instance \cite{zanotti2010hedging} and \cite{higgs2009modelling} apply CCC-GARCH models in electricity markets.

\subsection{ARX type models with bivariate jumps in the residuals with no constant mean and CCC-GARCH}

Our last model (ARX-BiJ-$\mu_d$-GARCH) includes CCC-GARCH structures in the continuous component of the model described in Equation \eqref{eq:bimean}. This is our most complex model, and it is a combination of the ARX-BiJ-$\mu_d$ and ARX-GARCH models:

\begin{align*}
&\bm{\mu}_{d} = \bm{\mu}_{0} + \bm{\mu}_{1}\bm{Y}_{d-1} \text{ and } \bm{\epsilon}_d = \bm{\epsilon}_{d,cont} + \bm{B}_d  \bm{\epsilon}_{d,jump} \\ 
& \text{with } 
\bm{\epsilon}_{d,cont} \sim \mathcal{N}_2(      - \bm{\Lambda} \bm{\mu}_d,\,\bm{\Sigma}_d), 
\text{diag}(\bm{B}_d) \sim \text{\textbf{Ber}}_2(\bm{P})  \text{ and }  
\bm{\epsilon}_{d,jump} \sim \mathcal{N}_2( \bm{\mu}_d ,\,\bm{\Gamma}), \\
&\sigma_{d,i}^2 = \alpha_{0,i} + \alpha_{1,i}\epsilon_{d-1,i}^2+\alpha_{2,i}\sigma_{d-1,i}^2 \ \text{for} \ i=1,2,\\
&\text{where} \ \ \bm{\Sigma}_d=\left( {\begin{array}{cc}
   \sigma_{d,1}^2 & \rho\sigma_{d,1}\sigma_{d,2} \\
   \rho\sigma_{d,1}\sigma_{d,2} & \sigma_{d,2}^2 \\
  \end{array} } \right)
\end{align*}
\noindent where all the components are assumed to be distributed as in the previous subsections. As it is the most complex model, it has the largest number of parameters to estimate. The model is able to capture all the aspects mentioned above.

\section{Estimation and Forecasting} \label{sec:estimation}

For the estimation we assume that there are $D$ observations available. We denote the resulting price vectors and regression matrix by $\mathbb{Y}_i = (Y_{1,i}, , \ldots, Y_{D,i})'$ and 
$\mathbb{X}_i =  (\bm{X}_{1,i}', , \ldots, \bm{X}_{D,i}')'$, corresponding to regression equation \eqref{eq:mean_lm}. 

To estimate the ARX model (Equation \eqref{eq:mean_lm}) we apply two different estimation methods: OLS\footnote{In order to avoid perfect collinearity in the OLS estimation we drop the interaction between Wednesday and the previous observations for both time series.} and elastic net. This gives us two different estimations and therefore two different forecasts, which we note as ARX-OLS and as ARX-enet, respectively.

Using  the OLS estimator the estimated values are:
$$
\widehat{\bm{\beta}}^{\text{OLS}}_i = \argmin_{\bm{\beta}\in\mathbb{R}^p} \left[ \| \mathbb{Y}_i - \mathbb{X}_i'\bm{\beta}  \|_2^2\right],
$$

The second estimation method applied to estimate Equation \eqref{eq:mean} is the elastic net, introduced by \cite{zou2005regularization}, which is very similar to OLS but has quadratic and linear penalties. However, in defining the elastic net estimator it is crucial to consider the corresponding scaled OLS problem. Hence, we introduce 
$\widetilde{\mathbb{Y}}_i $ and $ \widetilde{\mathbb{X}}_i$ as a scaled response vector and scaled regression matrix. We require them to be scaled in such a way that any column has a zero mean and standard deviation of 1.

Given the scaled OLS problem, the scaled elastic net estimator is given by the optimization problem
$$
 \widehat{\widetilde{\bm{\beta}} }_i^{\text{enet}} 
 = \argmin_{\bm{\beta}_\in\mathbb{R}^p} \left[ \| \widetilde{\mathbb{Y}}_i - \widetilde{\mathbb{X}}_i'\bm{\beta}||_2^2 + \lambda\left(\frac{1-\alpha}{2} ||\bm{\beta}||_2^2 + \alpha||\bm{\beta}||_1\right)\right],
$$
\noindent where $\lambda$ and $\alpha$ are tuning parameters that characterize the penalty term $\lambda\left(\frac{1-\alpha}{2} ||\bm{\beta}||_2^2 + \alpha||\bm{\beta}||_1\right)$.
 We receive the (unscaled) elastic net estimator $\widehat{\bm{\beta}}_i^{\text{enet}}$ simply by rescaling $\widehat{\widetilde{\bm{\beta}}}_i^{\text{enet}}$. 
If $\alpha=1$ the estimation method is equivalent to the lasso penalty developed by \cite{tibshirani1996regression}, and when $\alpha=0$ it is equivalent to the ridge penalty first introduced by \cite{hoerl1970ridge}.
The lasso estimator has the property of  sparsity, which means that for certain values of $\lambda$ the resulting solution sets irrelevant parameters to zero while keeping relevant parameters at non-zero.
The lasso estimation enjoys some popularity in electricity price forecasting: see \cite{ ziel2016forecasting, gaillard2016additive,steinert2019short, narajewski2019econometric}, amongst others.

The elastic net can be seen as an augmented data lasso shrinkage with some ridge elements. Like the lasso, the elastic net has automatic sparsity property for $\lambda > 0$.
In the lasso estimation we apply cross validation (CV) to select the tuning parameter $\lambda$. It takes into account the number of observations, the number of parameters, the variance and the correlation, according to \cite{hebiri2013correlations}. However, \cite{uniejewski2016automated} conclude that better forecasts are achieved when the elastic net method is consider, thus we incorporate a ridge penalty to the lasso estimation method. \cite{uniejewski2016automated} apply elastic net and lasso estimation methods (amongst others) in the electricity price forecasting context. Their findings suggest that $\alpha = 0.5$ is a good choice for applications, so we apply it in this paper as well. We choose $\lambda$ by 10-fold block-CV. In order to control weekly seasonal dependency structure we consider block-CV, the block length is 7 so that weekly seasonality is taken into account. This estimation method is often used in time series analysis, see. e.g. 
\cite{racine2000consistent}.

\begin{table}[ht]
\centering
\begin{tabular}{|r|r|ll||r|r|ll|}
  \hline
 Parameter& Name & off-peak & peak & Parameter& Name & off-peak & peak \\ 
  \hline
$\beta_{i,0} $&cons & $\cellcolor[rgb]{0.5,1,0.5} 1$ & $\cellcolor[rgb]{0.5,1,0.5} 1$               &$\beta_{i,3,3} $ &Wed                   & $\cellcolor[rgb]{1,0.659,0.5} 0.27$ & $\cellcolor[rgb]{1,0.518,0.5} 0.03$ \\  
$\beta_{i,1,1} $ &  $\text{AR}_{1,1}$ & $\cellcolor[rgb]{0.5,1,0.5} 1$ & $\cellcolor[rgb]{0.5,1,0.5} 1$     &$\beta_{i,4,3} $ &Thu                   & $\cellcolor[rgb]{1,0.585,0.5} 0.14$ & $\cellcolor[rgb]{0.951,0.849,0.5} 0.56$ \\  
$\beta_{i,2,1} $&  $\text{AR}_{2,1}$ & $\cellcolor[rgb]{1,0.714,0.5} 0.36$ & $\cellcolor[rgb]{0.909,0.891,0.5} 0.61$  & $\beta_{i,5,3} $&Fri                   & $\cellcolor[rgb]{1,0.535,0.5} 0.06$ & $\cellcolor[rgb]{0.997,0.803,0.5} 0.5$ \\ 
$\beta_{i,3,1} $&  $\text{AR}_{3,1}$ & $\cellcolor[rgb]{1,0.564,0.5} 0.11$ & $\cellcolor[rgb]{0.95,0.85,0.5} 0.56$    & $\beta_{i,6,3} $&Sat                   & $\cellcolor[rgb]{0.5,1,0.5} 1$ & $\cellcolor[rgb]{0.681,1,0.5} 0.85$ \\ 
$\beta_{i,4,1} $&  $\text{AR}_{4,1}$ & $\cellcolor[rgb]{1,0.776,0.5} 0.46$ & $\cellcolor[rgb]{0.821,0.979,0.5} 0.72$  & $\beta_{i,7,3} $&Sun                   & $\cellcolor[rgb]{0.5,1,0.5} 1$ & $\cellcolor[rgb]{0.5,1,0.5} 1$ \\ 
$\beta_{i,5,1} $&  $\text{AR}_{5,1}$ & $\cellcolor[rgb]{0.58,1,0.5} 0.93$ & $\cellcolor[rgb]{0.843,0.957,0.5} 0.7$   & $\beta_{i,1,4} $&Mon $\text{AR}_{1,1}$ & $\cellcolor[rgb]{0.981,0.819,0.5} 0.52$ & $\cellcolor[rgb]{0.676,1,0.5} 0.85$ \\ 
$\beta_{i,6,1} $&  $\text{AR}_{6,1}$ & $\cellcolor[rgb]{1,0.525,0.5} 0.04$ & $\cellcolor[rgb]{0.761,1,0.5} 0.78$     & $\beta_{i,2,4} $&Tue $\text{AR}_{1,1}$ & $\cellcolor[rgb]{1,0.698,0.5} 0.33$ & $\cellcolor[rgb]{0.972,0.828,0.5} 0.53$ \\ 
$\beta_{i,7,1} $&  $\text{AR}_{7,1}$ & $\cellcolor[rgb]{1,0.64,0.5} 0.23$ & $\cellcolor[rgb]{0.947,0.853,0.5} 0.57$   & $\beta_{i,3,4} $&Wed $\text{AR}_{1,1}$ & $\cellcolor[rgb]{1,0.77,0.5} 0.45$ & $\cellcolor[rgb]{1,0.766,0.5} 0.44$ \\ 
$\beta_{i,8,1} $&  $\text{AR}_{8,1}$ & $\cellcolor[rgb]{0.566,1,0.5} 0.95$ & $\cellcolor[rgb]{1,0.567,0.5} 0.11$      & $\beta_{i,4,4} $&Thu $\text{AR}_{1,1}$ & $\cellcolor[rgb]{1,0.532,0.5} 0.05$ & $\cellcolor[rgb]{1,0.527,0.5} 0.05$ \\  
$\beta_{i,1,2} $&  $\text{AR}_{1,2}$ & $\cellcolor[rgb]{0.5,1,0.5} 1$ & $\cellcolor[rgb]{0.5,1,0.5} 1$                & $\beta_{i,5,4} $&Fri $\text{AR}_{1,1}$ & $\cellcolor[rgb]{1,0.5,0.5} 0$ & $\cellcolor[rgb]{1,0.525,0.5} 0.04$ \\  
$\beta_{i,2,2} $&  $\text{AR}_{2,2}$ & $\cellcolor[rgb]{0.5,1,0.5} 1$ & $\cellcolor[rgb]{0.866,0.934,0.5} 0.67$      & $\beta_{i,6,4} $&Sat $\text{AR}_{1,1}$ & $\cellcolor[rgb]{1,0.756,0.5} 0.43$ & $\cellcolor[rgb]{0.869,0.931,0.5} 0.66$ \\ 
$\beta_{i,3,2} $&  $\text{AR}_{3,2}$ & $\cellcolor[rgb]{0.98,0.82,0.5} 0.53$ & $\cellcolor[rgb]{0.5,1,0.5} 1$         & $\beta_{i,7,4} $&Sun $\text{AR}_{1,1}$ & $\cellcolor[rgb]{0.92,0.88,0.5} 0.6$ & $\cellcolor[rgb]{1,0.73,0.5} 0.38$ \\  
$\beta_{i,4,2} $&  $\text{AR}_{4,2}$ & $\cellcolor[rgb]{0.684,1,0.5} 0.85$ & $\cellcolor[rgb]{0.855,0.945,0.5} 0.68$  & $\beta_{i,1,5} $&Mon $\text{AR}_{1,2}$ & $\cellcolor[rgb]{1,0.742,0.5} 0.4$ & $\cellcolor[rgb]{0.864,0.936,0.5} 0.67$ \\  
$\beta_{i,5,2} $&  $\text{AR}_{5,2}$ & $\cellcolor[rgb]{1,0.592,0.5} 0.15$ & $\cellcolor[rgb]{0.851,0.949,0.5} 0.69$  & $\beta_{i,2,5} $&Tue $\text{AR}_{1,2}$ & $\cellcolor[rgb]{1,0.796,0.5} 0.49$ & $\cellcolor[rgb]{1,0.786,0.5} 0.48$ \\  
$\beta_{i,6,2} $&  $\text{AR}_{6,2}$ & $\cellcolor[rgb]{0.602,1,0.5} 0.92$ & $\cellcolor[rgb]{0.5,1,0.5} 1$          & $\beta_{i,3,5} $&Wed $\text{AR}_{1,2}$ & $\cellcolor[rgb]{1,0.516,0.5} 0.03$ & $\cellcolor[rgb]{0.845,0.955,0.5} 0.69$ \\ 
$\beta_{i,7,2} $&  $\text{AR}_{7,2}$ & $\cellcolor[rgb]{0.633,1,0.5} 0.89$ & $\cellcolor[rgb]{0.5,1,0.5} 1$          & $\beta_{i,4,5} $&Thu $\text{AR}_{1,2}$ & $\cellcolor[rgb]{1,0.789,0.5} 0.48$ & $\cellcolor[rgb]{0.979,0.821,0.5} 0.53$ \\ 
$\beta_{i,8,2} $&  $\text{AR}_{8,2}$ & $\cellcolor[rgb]{1,0.573,0.5} 0.12$ & $\cellcolor[rgb]{0.564,1,0.5} 0.95$     & $\beta_{i,5,5} $&Fri $\text{AR}_{1,2}$ & $\cellcolor[rgb]{1,0.709,0.5} 0.35$ & $\cellcolor[rgb]{0.946,0.854,0.5} 0.57$ \\  
$\beta_{i,1,3} $&  Mon & $\cellcolor[rgb]{0.797,1,0.5} 0.75$ & $\cellcolor[rgb]{0.792,1,0.5} 0.76$                    & $\beta_{i,6,5} $&Sat $\text{AR}_{1,2}$ & $\cellcolor[rgb]{1,0.678,0.5} 0.3$ & $\cellcolor[rgb]{0.969,0.831,0.5} 0.54$ \\ 
$\beta_{i,2,3} $&  Tue & $\cellcolor[rgb]{1,0.621,0.5} 0.2$ & $\cellcolor[rgb]{0.933,0.867,0.5} 0.58$                & $\beta_{i,7,5} $&Sun $\text{AR}_{1,2}$ & $\cellcolor[rgb]{1,0.768,0.5} 0.45$ & $\cellcolor[rgb]{0.945,0.855,0.5} 0.57$ \\  
   \hline
\end{tabular}
\caption{Percentage of times variables are included in elastic net estimation.}
\label{tab:enet}
\end{table}

In Table \ref{tab:enet} we show the percentage of times each one of the estimated parameters in Equation \eqref{eq:mean} is not equal to 0, i.e. the percentage of times parameters are included in the model. As can be observed, the autoregressive components are relevant either for the peak series or for the off-peak series. The day of the week dummies (mainly weekend effects) are relevant as well. Regarding the iteration between the day of the week and previous observation, the number of times estimated parameters are included in the models decreases.

We compare the results for the two estimation techniques. The elastic net provides better forecasting results, so in estimating the second step the residuals are calculated via the elastic net. 



\begin{figure}[h!]
\resizebox{0.5\textwidth}{!}{
 \subfloat[2016-01-01]{\includegraphics{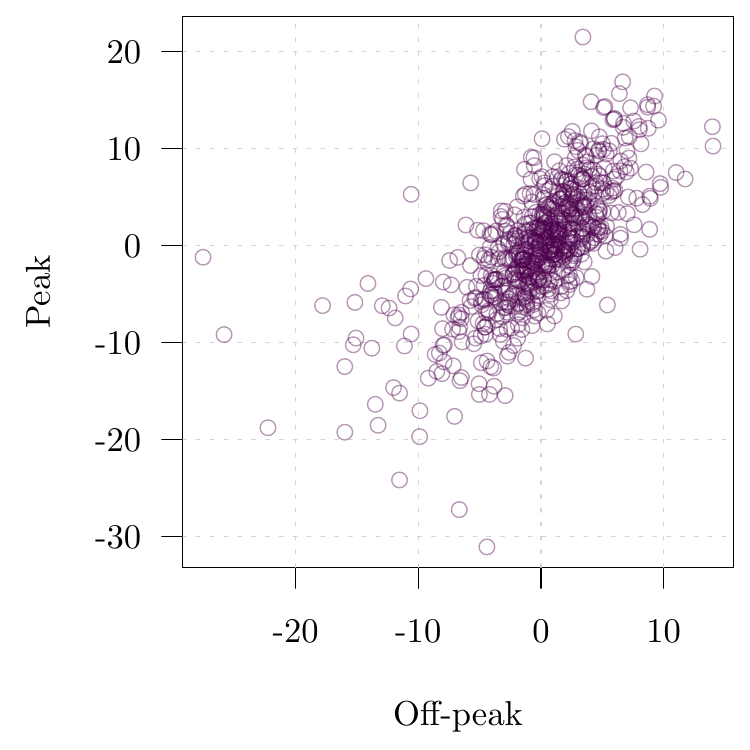}}}\hfill
\resizebox{0.5\textwidth}{!}{
 \subfloat[2017-12-31]{\includegraphics{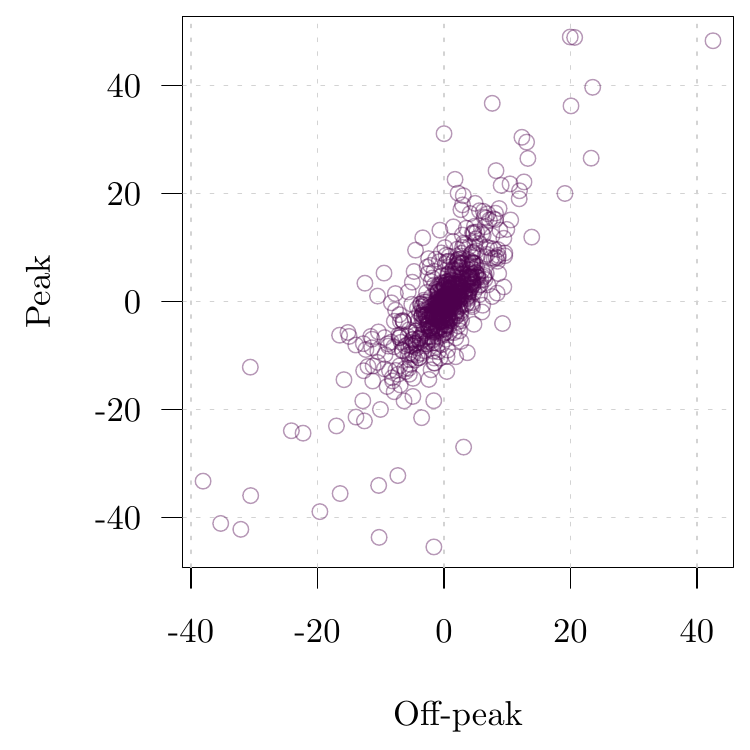}}}\hfill
\caption{Scatter plot of the residuals.}
\label{fig:resscatter}
\end{figure}

In Figure \ref{fig:resscatter}, the scatter plot of the residuals for two different days are shown, these are the residuals from the first of the rolling windows and the last one. We would like to point out firstly that the two scatter plots are very much alike which means that the ARX model performs similarly in both cases. Furthermore, the complex correlation structure is also observed, and is similar to the one in Figure \ref{fig:hist}. The density of the scatter plot may be observed since the darker the color, the greater the number of points clustered.  

In the second step residuals are estimated by maximum likelihood\footnote{In the ARX-IJ model, $\rho$ is estimated in a third stage as $\rho=\text{Cor}(\epsilon_{d,1},\epsilon_{d,2})$.} using the Broyden-Fletcher-Goldfarb-Shanno (BFGS) algorithm. This algorithm is a quasi-Newton method for non-linear optimization. In order to apply the maximum likelihood estimation we need a log-likelihood function which changes depending on the structure of the residuals assumed. That log-likelihood function is based on the assumed distribution for each of the different models explained in Section \ref{sec:model}. The maximum likelihood estimation is initialized as follows: 
\begin{itemize}
\item In the ARX-IJ model, as both residuals series are independently estimated, the starting values are taken as $\sigma , \gamma =$ standard deviation of the time series, $\mu =1$ and $\lambda=0.01$ for each one of the time series.
\item In the ARX-biJ model, the parameters $\sigma_1,\sigma_2,\mu_1,\mu_2,\gamma_1$ and $\gamma_2$ are started in the estimated values of the model ARX-IJ. Additionally, $p_{1,0}=0.01, p_{0,1}=0.01, p_{1,1}=0.001, \rho=0.01$ and $\varrho=0.01$.  
\item In the ARX-biJ-$\mu_d$ model, the initial values are set to $\mu_{1,1} ,\mu_{1,2} = 0.01$ and the other values are taken from the estimated parameters of the ARX-biJ model.
\item In the ARX-GARCH model, parameters are initialized at values $\rho,\alpha_{1,1},\alpha_{2,1},\alpha_{1,2},\alpha_{2,2}=0.01$ and $\alpha_{0,1}, \alpha_{0,2}$ are the standard deviations of the respective residuals series.
\item Finally, in the ARX-biJ-$\mu_d$-GARCH model, the paramters  $\alpha_{1,1},\alpha_{2,1},\alpha_{1,2},\alpha_{2,2}$ are again set to 0.01, and the other initial values are taken from the model  ARX-biJ-$\mu_d$.
\end{itemize}
The above procedure to calculate the initial values ensures that whenever we include a new component, if the estimated parameters of this component are not zero, then the component improves the log likelihood function performance.

Once all the parameters are estimated, off-peak and peak time series are simulated. In our case we predict $H=7$ horizons and for each of horizon $M=16000$ paths are simulated. All the simulations (sometimes called ensemble) can be seen as multivariate probabilistic forecasts as they approximate well the underlying distribution of the forecasts. All 
relevant properties can be derived from these paths. It is possible to analyze only the marginal properties of each predicted horizon. 
The estimation and simulation process is repeated $N=731$ times, as mentioned in Section \ref{sec:data}, via a rolling window. The first estimation is made using the first two years of the data, then the next $H=7$ days are predicted with $M$ paths in each horizon. Then, the estimation sample is shifted one day forward and the process is repeated.

\section{Evaluation criteria} \label{sec:evaluation}

In this section we introduce the evaluation criteria used to assess probabilistic forecasting. In total we use four different criteria: MAE, MSE, PB, and ES. We first explain the scores that we use, then briefly introduce the DM test used to test whether differences in forecasting performance are significant or not, taking the models in pairs. To compute the DM test it is necessary to define a loss function. We therefore introduce each criterion with the corresponding loss function. As mentioned in Section \ref{sec:estimation} we forecast seven horizons, and all the four criteria are independently evaluated for each of the horizons.

For evaluating the point forecasts, we consider the popular MAE and MSE measures. The MAE is a strictly proper forecasting criterion for the median and the MSE a strictly proper evaluation criterion for the mean, with "strictly proper" here meaning that only 
the perfect model minimizes the corresponding criterion. 
Therefore we define $\widehat{Y}_{d,i}^{\text{med}}$ as a median forecast and $\widehat{Y}_{d,i}$ mean for day $d$ and volatility series $i$ derived from the sample counterparts of the $M$ simulated paths.
The MAE and MSE are defined using the absolute error (AE) and the squared error (SE).
Thus, with 
\begin{align}
 \text{AE}_{d,i} &= |Y_{d,i}-\widehat{Y}^{\text{med}}_{d,i}| \label{eq:med} \\ 
\text{SE}_{d,i} &= \left(Y_{d,i}-\widehat{Y}_{d,i}\right)^2 
\end{align}
\noindent equation \eqref{eq:med} is often used in the literature for mean forecasts but it is not proper from a statistical point of view. It is proper when we have symmetry in the sample as in this case median is equal to mean. We define,
\begin{align}
\text{MAE}_i = \frac{1}{N}\sum_{d=1}^N \text{AE}_{d,i}  \\ 
 \text{MSE}_i = \frac{1}{N}\sum_{d=1}^N \text{SE}_{d,i} 
\end{align}
for $ i=1,2$.
Hence, we can evaluate the point forecasts for the off-peak and peak price separately.

The two criteria introduced above are the most widely used in the literature, but we are more interested in the marginal properties of the models, and so we use the PB.

The PB measures the distance for each quantile. 
Therefore, as for the median and mean forecasts,  we define $\widehat{Y}_{d,q,i}$ as a forecast for the $q$-quantile day $d$ and time series $i$. 
We get these quantile forecasts by taking the sample quantile of our $M$ simulated paths.
The pinball loss function is computed as follows,
\begin{equation}
PB_{d,q,i}=\begin{cases} (1-q) (\widehat{Y}_{d,q,i}-Y_{d,i}), & \mbox{if }  \widehat{Y}_{d,q,i}\geq Y_{d,i}\\ 
q(Y_{d,i}-\widehat{Y}_{d,q,i}), & \mbox{if } \widehat{Y}_{d,q,i}< Y_{d,i} \end{cases}   \ \ \text{for} \ \ i=1,2 ,
\label{eq:PB}
\end{equation}

\noindent where $q\in \bm{Q}$ is a quantile, in our case $\bm{Q}= \{Q_q\}_{q\in \{1,\ldots, K\}} = \{ 0.01,0.02,\ldots,0.99 \}$ with $K=99$. $\widehat{Y}_{d,q,i}$ stands for the estimated price of quantile $q$ on day $d$ and in time series $i$ and $Y_{d,i}$ is the observed value at day $d$ and time series $i$. To calculate the PB of quantile $q$, series $i$ and N days, we proceed as follows:
$$
PB_{q,i} =\frac{1}{N}\sum_{d=1}^N PB_{d,q,i}  \ \ \text{for} \ \ i=1,2 \ \ \text{and} \ \ q\in Q,
$$
Thus, the PB for $N$ days is computed by averaging across the $\bm{Q}$ quantiles,
$$
PB_{i} =\frac{1}{K}\sum_{q=1}^K PB_{Q_q,i}  \ \ \text{for} \ \ i=1,2 \ \ \text{and} \ \ q\in Q,
$$
\noindent where $K$ is the number of quantiles. Note that if the distance in the quantile grid $\bm{Q}$ converges to $0$ then the PB converges to the probabilistic forecasting
evaluation measure CRPS (continuous ranked probability score), which is strictly proper with respect to the  (marginal) distribution of $Y_i$. For further information on the PB score, see \cite{steinwart2011estimating}.

When applying PB we may observe all the marginal properties of the forecasts, given that we may observe the forecasting performance of the different models in each quantile. Therefore, the first criterion introduced in this section is merely a special case of PB when $q=0.5$. With this score we can compare how the different models capture spikes, as we can analyze the behavior in the tails.

The last criterion that we use is the ES, which is a generalization of the CRPS. The ES is the only score that takes into account the dependency structures. This score is applied to all the variables at the same time in order to take in the correlation. In both cases, in the modeling and evaluating we pay close attention to the dependency structure as this is the main contribution of our paper. The loss function of the ES is computed as follows:

\begin{align}
\text{ES}_d=& \text{ED}_d - \frac{1}{2}\text{EI}_d \label{eq:ES}\\
\text{ED}_d =& \frac{1}{M}\sum_{m=1}^M \left\Vert \bm{Y}_d^{[m]}- \bm{Y}_d \right\Vert_2 \nonumber \\
\text{EI}_d =& \frac{1}{M}\sum_{m=1}^M \left\Vert \bm{Y}_d^{[m]}- \bm{Y}_d^{[m+1]} \right\Vert_2 \ \ \text{where} \ \  \bm{Y}_d^{[M+1]}= \bm{Y}_d^{[1]} \nonumber
\end{align}

\noindent
$\bm{Y}_d^{[m]}$ for $m=1,\ldots,M$ is the predicted $m^{th}$ path of the multivariate data for day $d$. Our estimator for the ES is based on \cite{gneiting2007strictly} and is divided in two parts: the ED and EI. The ED measures the mean Euclidean distance from each one of the paths to the observed value, so this part is measuring marginal properties. On the other hand, the EI measures within the path dependencies, i.e. how well the path are aligned with each other. The EI appears with a negative sign which means we are interested in spread amongst trajectories. Equation \eqref{eq:ES} is the ES for day $d$, thus the ES for N days is calculated as,
$$
\text{ES} = \frac{1}{N} \sum_{d=1}^N \text{ES}_d.
$$

The next step is to check whether or not the differences between the forecasting performances with each criterion are significantly different from zero. For the significance test we use the DM test, which compares models in pairs. As mentioned above, we need a loss function such as the ones written above to apply the DM test. Let $L_d$ denote the loss function of a certain model; the loss differential between models $\mathbb{A}$ and $\mathbb{B}$ is defined as $\delta_{d,\mathbb{A},\mathbb{B}}=L_{d,\mathbb{A}}-L_{d,\mathbb{B}}$. The only required assumption is for the loss differential to be covariance stationary. To apply the DM test we compute

$$
\frac{\widetilde{\delta}_{\mathbb{A},\mathbb{B}}}{\sigma_{\widetilde{\delta}_{\mathbb{A},\mathbb{B}}}} \sim \mathcal{N}_1 (0,1)
$$

\noindent where $\widetilde{\delta}_{\mathbb{A},\mathbb{B}}=\frac{1}{N}\sum_{d=1}^N \delta_{d,\mathbb{A},\mathbb{B}}$ and $\sigma_{\widetilde{\delta}_{\mathbb{A},\mathbb{B}}}$ is the 
standard error which we estimate by the corresponding sample counterpart. For further information on the DM test, see \cite{diebold2002comparing} and \cite{diebold2015comparing}.

\section{Results} \label{sec:results}

In this section we assess the forecasting performance of each model, using the criteria introduced in Section \ref{sec:evaluation}. Then, the DM test is applied to check whether the differences between the models in pairs are significant or not\footnote{DM test results for all criteria and all horizons are available upon request.} for each criterion. The forecasting horizon (H) is 7, which means that for each rolling window we get the forecasts for the following 7 days. In order to show what the different trajectories look like, we present trajectories for the ARX-enet and ARX-BiJ-$\mu_d$ models at the end of the Section.




 In the following graphs we show how each one of the models performs using all the four criteria from Section \ref{sec:evaluation}. According to all the evaluation criteria the optimal scores zero, so the lower the scored value the better the forecasting performance. In all four criteria, the relative performances compared to the ARX-OLS model are shown, i.e. the performance of each one of the models is divided by the ARX-OLS values for each one of the horizons Score(model)/Score(ARX-OLS).

\begin{figure}[h!]
\centering
\resizebox{0.495\textwidth}{!}{
 \subfloat[Relative off-peak MAE criterion]{\includegraphics{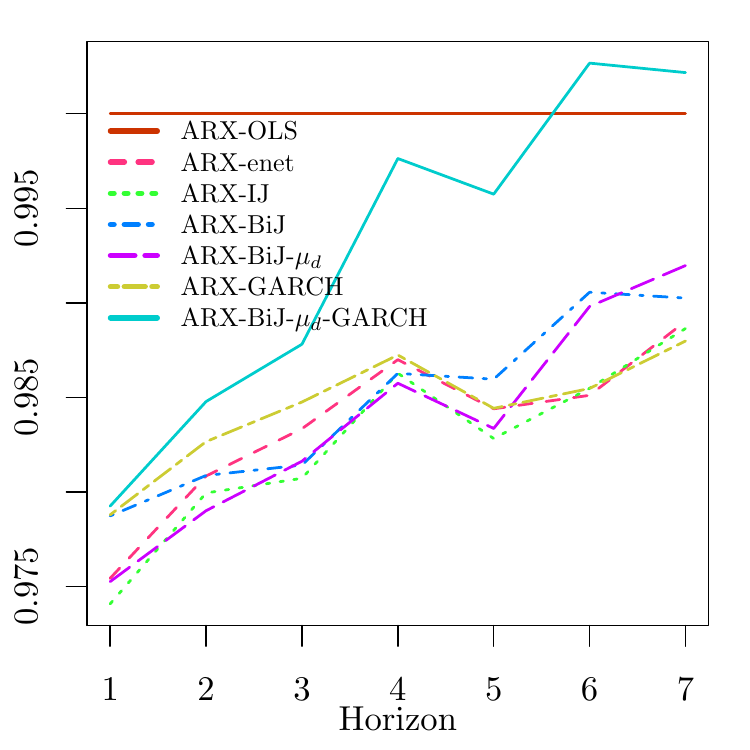}}}
 \resizebox{0.495\textwidth}{!}{
 \subfloat[Relative peak MAE criterion]{\includegraphics{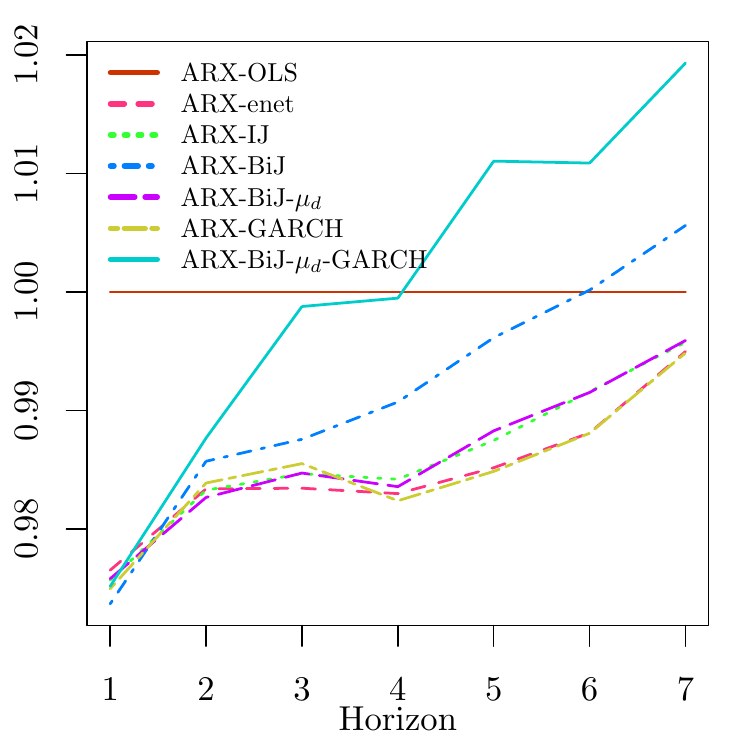}}}\\
  \resizebox{0.495\textwidth}{!}{
 \subfloat[ARX-OLS MAE criterion]{\includegraphics{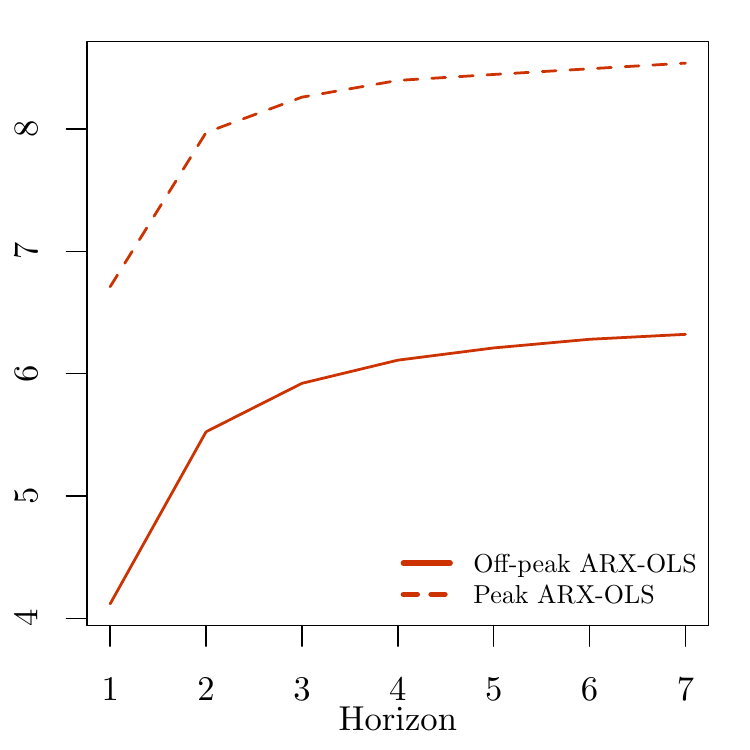}}}
\caption{MAE criterion performance.}
\label{fig:mae}
\end{figure}

Figure \ref{fig:mae} shows the results for the MAE criterion for the two time series. As explained above, the MAE criterion is a special case of the PB that compares performance on the median. As observed in Figure \ref{fig:mae}, the differences between the models are not too big: the only clear result is that the performance of the ARX-OLS model is significantly poorer. Overall, results for the ARX-GARCH and ARX-IJ can be said to be slightly better but, the forecasts are not significantly better according to the DM test. It is important to underline that a comparison between the OLS and elastic net estimation methods reveals that the latter gives significantly better forecasting results. 

\begin{figure}[h!]
\centering
\resizebox{0.495\textwidth}{!}{
 \subfloat[Relative off-peak MSE criterion]{\includegraphics{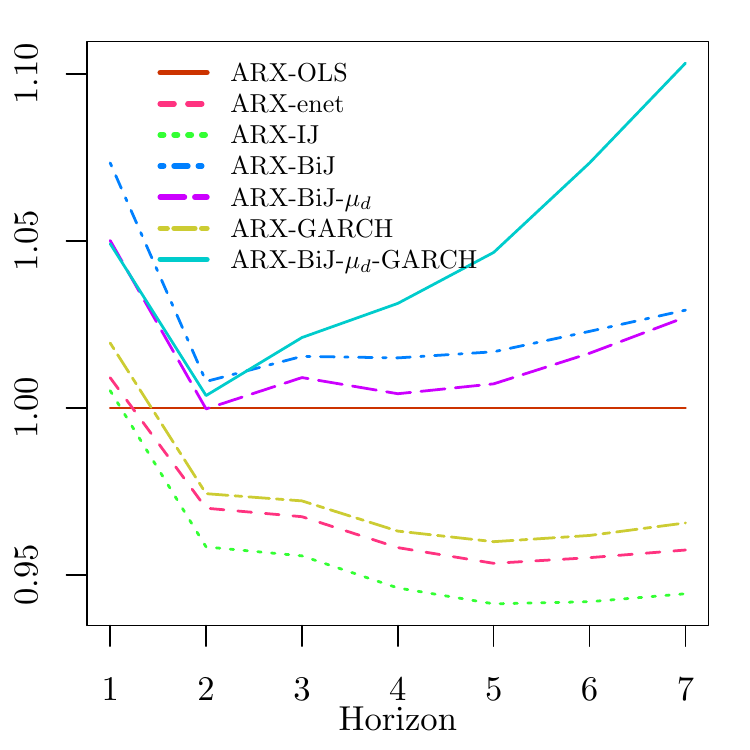}}}
 \resizebox{0.495\textwidth}{!}{
 \subfloat[Relative peak MSE criterion]{\includegraphics{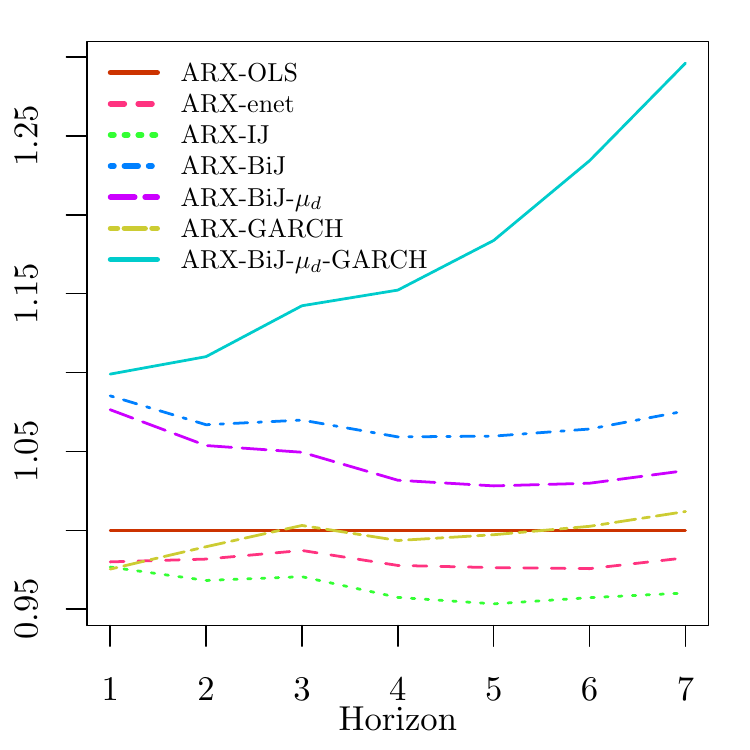}}}\\
  \resizebox{0.495\textwidth}{!}{
 \subfloat[ARX-OLS MSE criterion]{\includegraphics{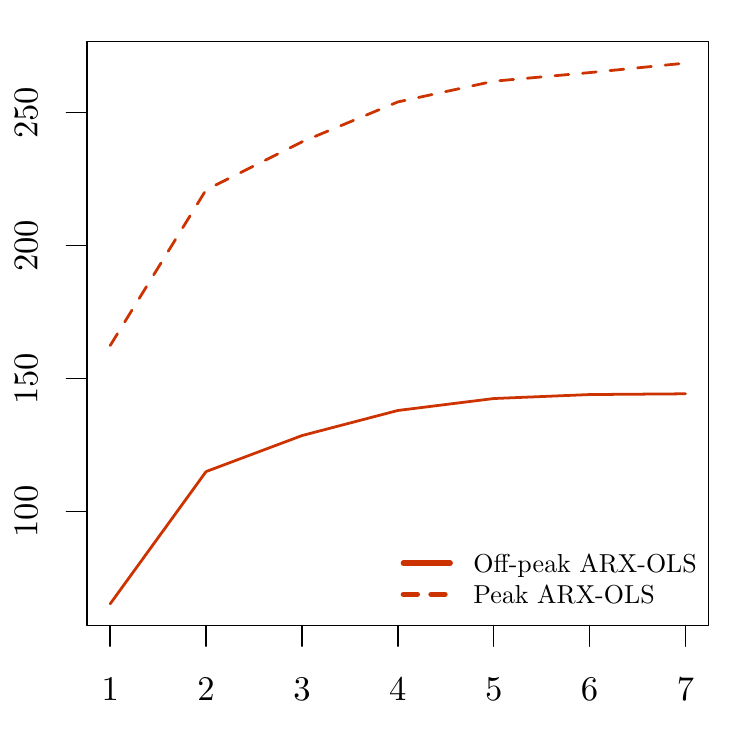}}}
\caption{MSE criterion performance.}
\label{fig:mse}
\end{figure}

Figure \ref{fig:mse} shows the forecasting performance at the mean. A look at Figure \ref{fig:mse} and the DM test suggests that bivariate jump structures are not effective in capturing mean behavior, as the forecasting performance is poor according to the MSE criterion. At the same time, the Figure shows that the forecasts of the ARX-IJ are better than other models and the DM test confirms that the differences are significant, with the exception of the first horizon, where the difference between ARX-IJ and ARX-enet is not significant for either of the time series. The latter means that for the mean forecast, it is important to introduce jump structures; but dependency structures are not highly relevant. The superiority of the ARX-IJ forecasting performance becomes greater when the horizons are increased. According to this criterion, elastic net forecasts are significantly better than the OLS forecasts when a simple ARX model is simulated. 

\begin{figure}[h!]
\centering
\resizebox{0.495\textwidth}{!}{
 \subfloat[Relative off-peak PB score]{\includegraphics{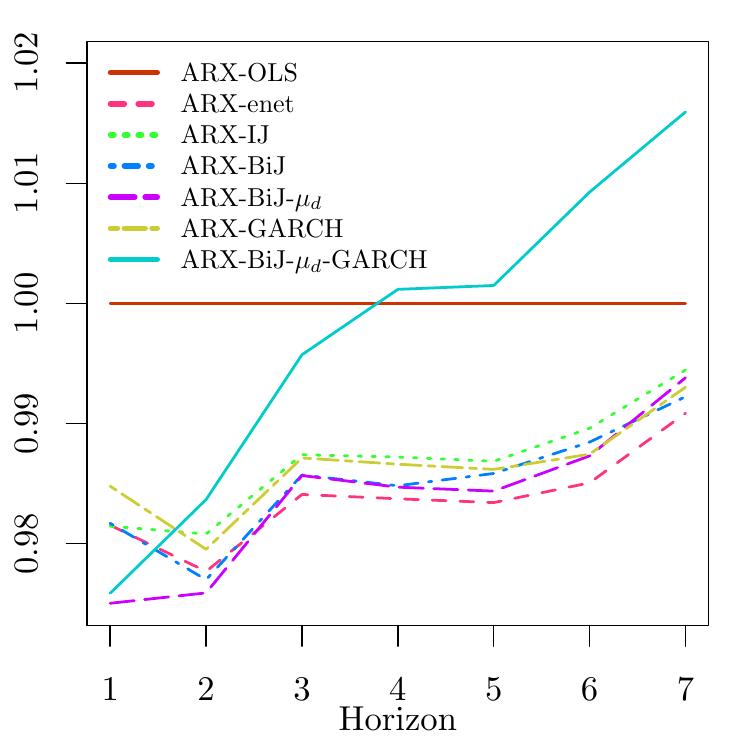}}}
 \resizebox{0.495\textwidth}{!}{
 \subfloat[Relative peak PB score]{\includegraphics{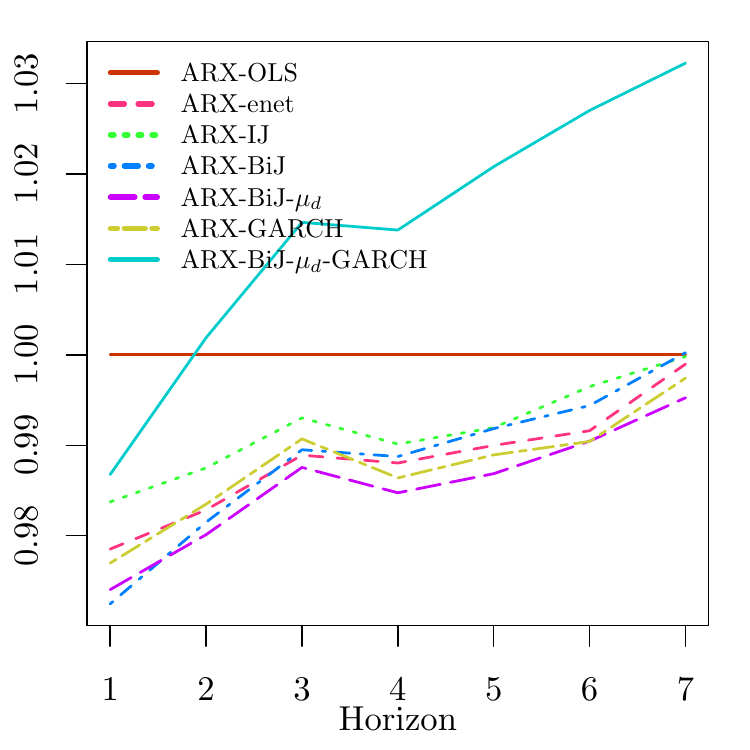}}}\\
  \resizebox{0.495\textwidth}{!}{
 \subfloat[ARX-OLS PB score]{\includegraphics{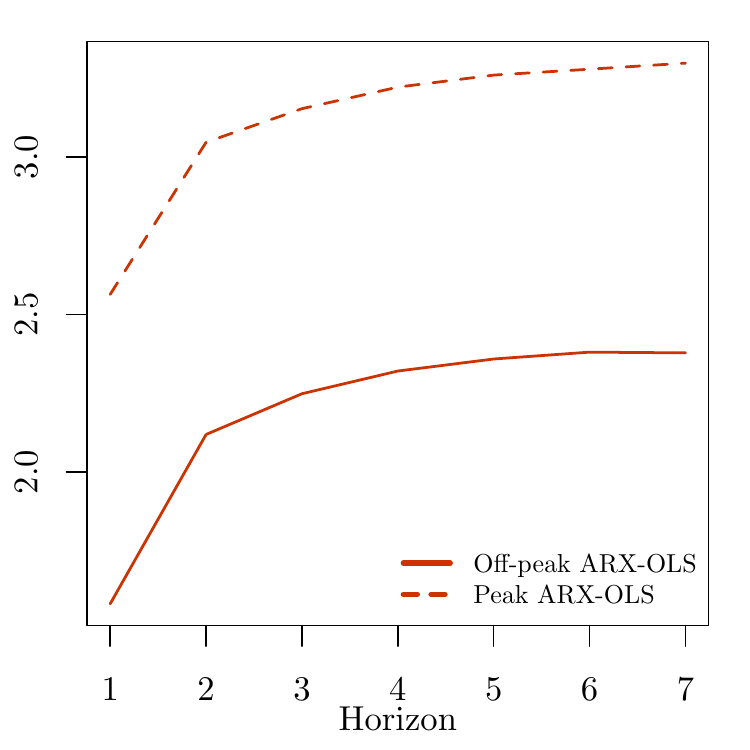}}}
\caption{PB score performance.}
\label{fig:PB}
\end{figure}

As shown in the previous section, the PB takes into account the whole distribution of the forecast paths quantile by quantile. As shown in Figure \ref{fig:PB}, in the case of the peak time series the ARX-BiJ-$\mu_{d}$ models forecast outperforms the other models except in one case ( in H2 the PB of the ARX-BiJ model is lower). On the other hand, in the off-peak seriesfor the first horizons, the ARX-BiJ forecasts are seen to be the best, but after the $6^{\text{th}}$ horizon the ARX-enet has the best forecasting performance. This is curious because the ARX-enet does not take into account heavy tails and a model with jumps would be expected to capture tail behavior more efficiently. However, it must be underlined that the differences between these models are not significant according to the DM test. The DM test only concludes that in both time series the forecasting performance is significantly poorer for the ARX-OLS and the ARX-BiJ-$\mu_{d}$-GARCH models. As with the MAE and MSE criteria, with the PB score the elastic net estimation method provides better forecasts than the OLS .

\begin{figure}[h!]
\subfloat[off-peak quantile performance]{\includegraphics{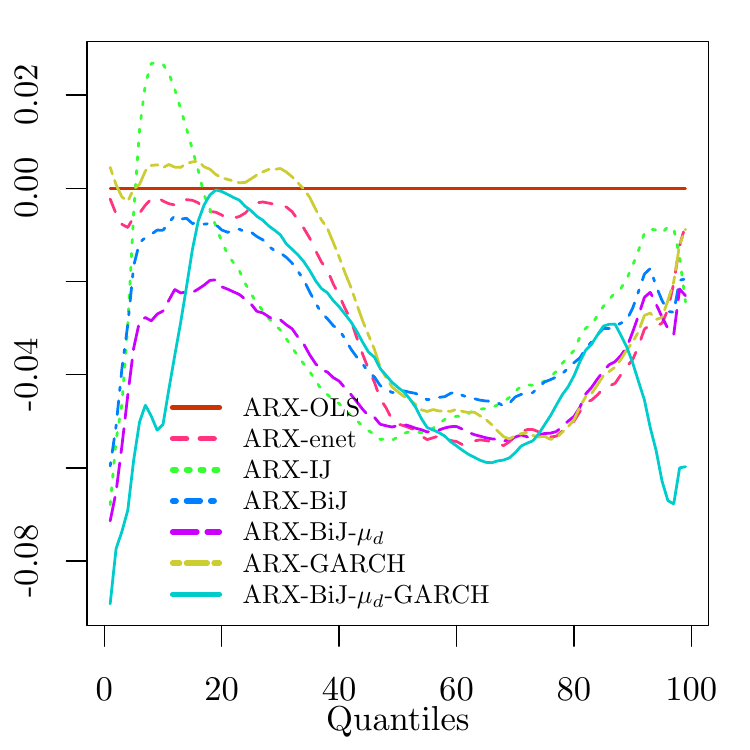}}
\subfloat[peak quantile performance]{\includegraphics{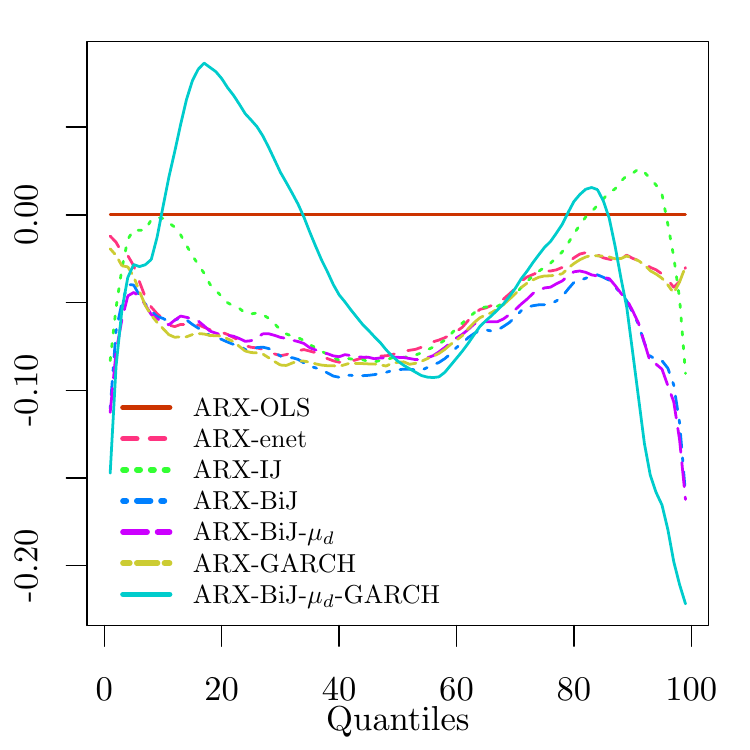}}
\caption{Performance on each quantile for first horizon}
\label{fig:quantile}
\end{figure}

Figure \ref{fig:quantile} shows $PB(model)-PB(OLS)$ for the first forecast horizon quantile by quantile, that is all the models are compared to the ARX-OLS. Focusing on the model ARX-BiJ-$\mu_{d}$-GARCH, which is the most complex model, it may be observed that it is the best model in the first and the last quantiles, but in the middle quantiles it has the worst forecasting performance. This is even more pronounced for the peak time series. The behavior of the forecasts with ARX-BiJ-$\mu_{d}$ and ARX-BiJ models is similar in both time series. In the off-peak time series, we observe that in the middle quantiles the forecasting performance is the best for the ARX-enet and the ARX-GARCH. Furthermore, it is surprising that the ARX-IJ performs so poorly in the highest and lowest quantiles because unlike ARX-OLS, ARX-enet and ARX-GARCH models, the ARX-IJ model considers spikes which captures behavior on the tails.

So far we have distinguished between the two time series because dependencies are not taken into account in the criteria mentioned above. It may be observed for the three previous criteria that the off-peak time series has a lower error term, which means that forecasts are more accurate according to all three criteria. This fact is expected as the off-peak time series is less volatile and therefore easier to predict. The ES takes into account dependency structures, which are the key feature of this paper.

We start analyzing the ES performance by showing the two parts defined on Equation \eqref{eq:ES}. The ED measures the euclidean distance, and only takes into account marginal properties. This measure is similar to the MSE measure but taking the two time series at the same time. Thus, as in Figure \ref{fig:mse} there are no big differences  between the two times series performance, we are expecting a similar performance with the ED measure. The EI measures within the path dependency, i.e. how well the trajectories are aligned with each other. 
 
\begin{figure}[h!]
\centering
\resizebox{0.495\textwidth}{!}{
\subfloat[Relative ED score]{\includegraphics{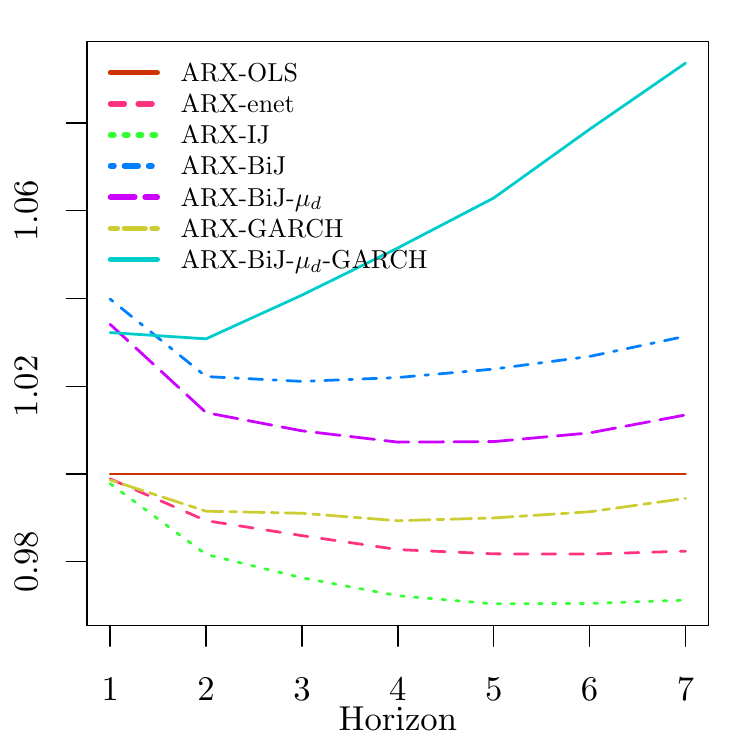}}}
\resizebox{0.495\textwidth}{!}{
\subfloat[Relative EI score]{\includegraphics{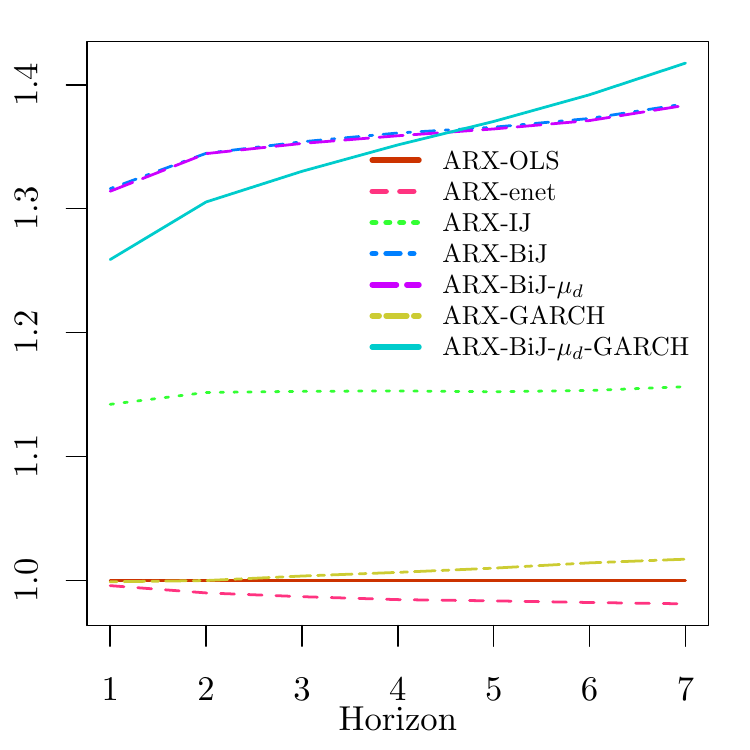}}}\\
\resizebox{0.495\textwidth}{!}{
\subfloat[ARX-OLS ED and EI scores]{\includegraphics{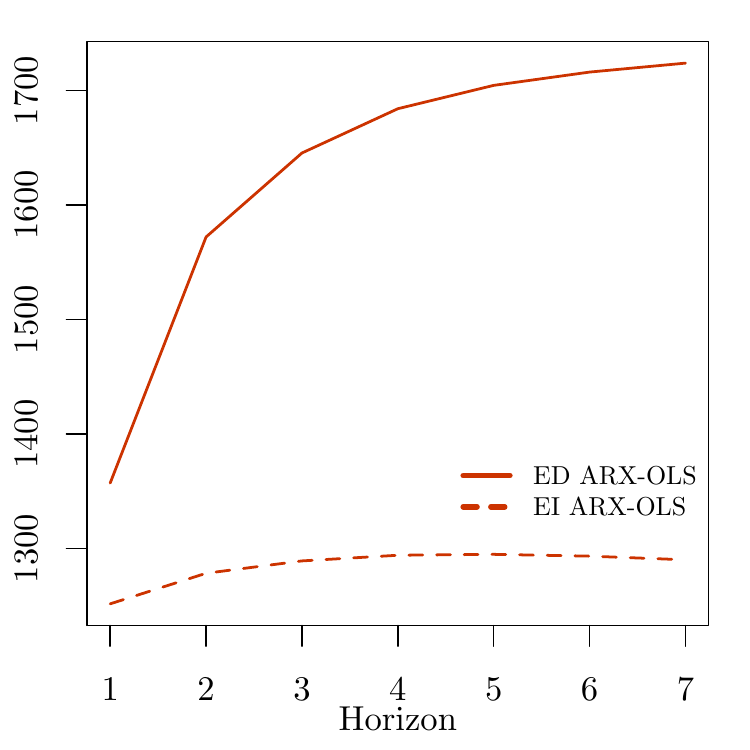}}}
\caption{ED and EI performance.}
\label{fig:EDEI}
\end{figure}

Figure \ref{fig:EDEI} shows the forecasting performance for the two parts of the ES. According to the ES criterion a model has a better forecasting performance when the estimated value of the score is lower. Thus, on the ED lower values are needed while on the EI higher ones are better. Focusing on the ED, poorer performance of the models with bivariate jumps is observed (similar to the MSE criterion), and the ARX-IJ model performs the best. Regarding the EI, it is shown that including spikes in the models increases the spread of the trajectories considerably, thus increasing the estimated value of the EI. Moreover, incorporating bivariate jumps increases the spread even more. The differences on the ED are lower than the differences on the EI, and even if the EI value is multiplied by 0.5 the difference is still bigger. Consequently, models containing bivariate jumps perform better on the ES.

\begin{figure}[h!]
\resizebox{0.49\textwidth}{!}{
\subfloat[Relative ES performance]{\includegraphics{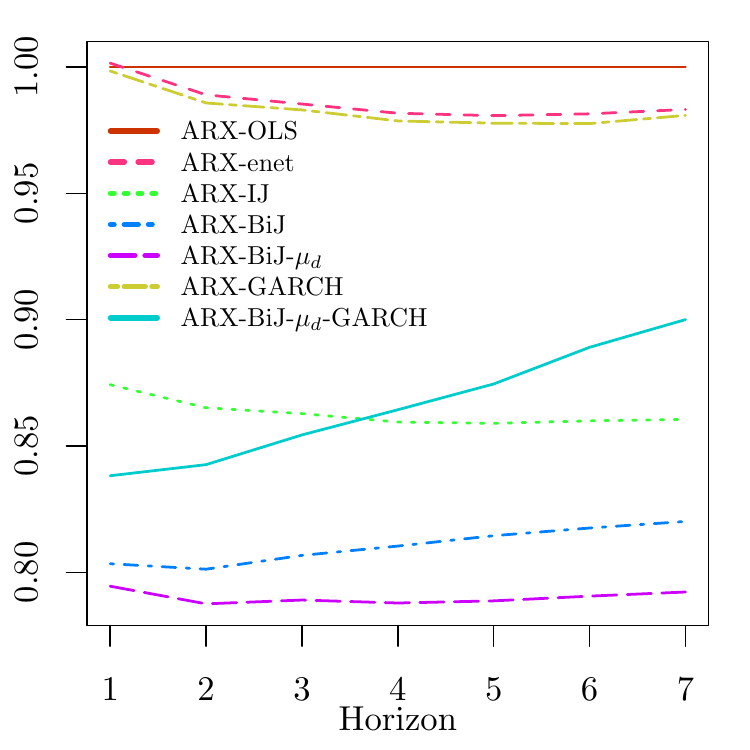}\label{fig:ES1}}}
\resizebox{0.49\textwidth}{!}{
\subfloat[ES score ARX-OLS model]{\includegraphics{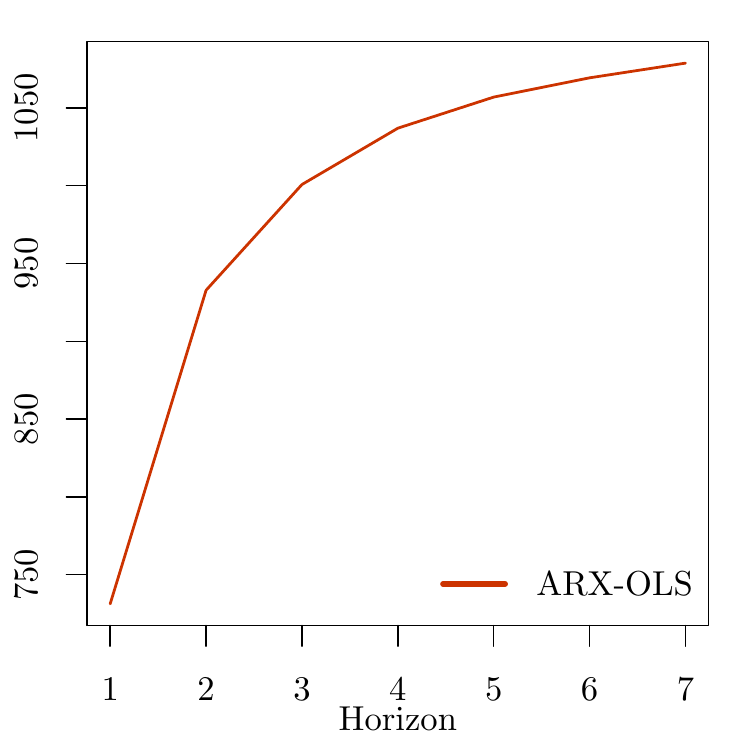}\label{fig:ES2}}}
\caption{Energy score performance.}
\label{fig:ES}
\end{figure}

Figure \ref{fig:ES} shows the ES for our seven models and our seven horizons. In Figure \ref{fig:ES1} the relative performance against the ARX-OLS model is shown and in Figure \ref{fig:ES2} the shape of the ARX-OLS model is depicted. It is observed that the best forecasts are made by the ARX-BiJ-$\mu_d$, which makes sense given that it incorporates a complex dependency structure. One would expect the ARX-BiJ-$\mu_d$-GARCH to capture all features of the time series better, but it may the case that there are too many parameters to estimate: with our starting values we have hit a local maximum\footnote{We have tried different starting values but the ones used in the estimation have the best results. Other starting values could be used but this would greatly increase a lot the computation time. As 731 estimations are made in total, increasing the number of starting values would make estimation unfeasible.}. Another reason might be structural breaks in the dataset, leading to poor forecasting performance. As already mentioned in the previous paragraph, in Figure \ref{fig:ES} is shown that that the models with bivariate jumps generally provide better forecasts with this criterion, which means that the dependency structures are properly included by considering bivariate jumps. This is an important result as our objective is to efficiently capture the complex correlation structure between our price series.

\begin{table}[h!]
\centering
\resizebox{\textwidth}{!}{
\begin{tabular}{rlllllll}
  \hline
Models & M1 & M2 & M3 & M4 & M5 & M6 & M7 \\ 
  \hline
M1 &  & $\cellcolor[rgb]{1,0.706,0.5} \underset{(0.381)}{-0.3}$ & $\cellcolor[rgb]{1,0.5,0.5} \underset{(>0.999)}{23.48}$ & $\cellcolor[rgb]{1,0.5,0.5} \underset{(>0.999)}{32.87}$ & $\cellcolor[rgb]{1,0.5,0.5} \underset{(>0.999)}{36.87}$ & $\cellcolor[rgb]{1,0.628,0.5} \underset{(0.615)}{0.29}$ & $\cellcolor[rgb]{1,0.5,0.5} \underset{(>0.999)}{24.01}$ \\ 
  M2 & $\cellcolor[rgb]{1,0.627,0.5} \underset{(0.619)}{0.3}$ &  & $\cellcolor[rgb]{1,0.5,0.5} \underset{(>0.999)}{52.6}$ & $\cellcolor[rgb]{1,0.5,0.5} \underset{(>0.999)}{65.94}$ & $\cellcolor[rgb]{1,0.5,0.5} \underset{(>0.999)}{71.15}$ & $\cellcolor[rgb]{1,0.506,0.5} \underset{(0.983)}{2.11}$ & $\cellcolor[rgb]{1,0.5,0.5} \underset{(>0.999)}{36.17}$ \\ 
  M3 & $\cellcolor[rgb]{0.5,1,0.5} \underset{(<0.001)}{-23.48}$ & $\cellcolor[rgb]{0.5,1,0.5} \underset{(<0.001)}{-52.6}$ &  & $\cellcolor[rgb]{1,0.5,0.5} \underset{(>0.999)}{36.5}$ & $\cellcolor[rgb]{1,0.5,0.5} \underset{(>0.999)}{40.64}$ & $\cellcolor[rgb]{0.5,1,0.5} \underset{(<0.001)}{-49.06}$ & $\cellcolor[rgb]{1,0.5,0.5} \underset{(>0.999)}{9.07}$ \\ 
  M4 & $\cellcolor[rgb]{0.5,1,0.5} \underset{(<0.001)}{-32.87}$ & $\cellcolor[rgb]{0.5,1,0.5} \underset{(<0.001)}{-65.94}$ & $\cellcolor[rgb]{0.5,1,0.5} \underset{(<0.001)}{-36.5}$ &  & $\cellcolor[rgb]{1,0.5,0.5} \underset{(>0.999)}{5.97}$ & $\cellcolor[rgb]{0.5,1,0.5} \underset{(<0.001)}{-71.21}$ & $\cellcolor[rgb]{0.5,1,0.5} \underset{(<0.001)}{-9.42}$ \\ 
  M5 & $\cellcolor[rgb]{0.5,1,0.5} \underset{(<0.001)}{-36.87}$ & $\cellcolor[rgb]{0.5,1,0.5} \underset{(<0.001)}{-71.15}$ & $\cellcolor[rgb]{0.5,1,0.5} \underset{(<0.001)}{-40.64}$ & $\cellcolor[rgb]{0.5,1,0.5} \underset{(<0.001)}{-5.97}$ &  & $\cellcolor[rgb]{0.5,1,0.5} \underset{(<0.001)}{-64.69}$ & $\cellcolor[rgb]{0.5,1,0.5} \underset{(<0.001)}{-12.14}$ \\ 
  M6 & $\cellcolor[rgb]{1,0.705,0.5} \underset{(0.385)}{-0.29}$ & $\cellcolor[rgb]{0.837,1,0.5} \underset{(0.017)}{-2.11}$ & $\cellcolor[rgb]{1,0.5,0.5} \underset{(>0.999)}{49.06}$ & $\cellcolor[rgb]{1,0.5,0.5} \underset{(>0.999)}{71.21}$ & $\cellcolor[rgb]{1,0.5,0.5} \underset{(>0.999)}{64.69}$ &  & $\cellcolor[rgb]{1,0.5,0.5} \underset{(>0.999)}{36.1}$ \\ 
  M7 & $\cellcolor[rgb]{0.5,1,0.5} \underset{(<0.001)}{-24.01}$ & $\cellcolor[rgb]{0.5,1,0.5} \underset{(<0.001)}{-36.17}$ & $\cellcolor[rgb]{0.5,1,0.5} \underset{(<0.001)}{-9.07}$ & $\cellcolor[rgb]{1,0.5,0.5} \underset{(>0.999)}{9.42}$ & $\cellcolor[rgb]{1,0.5,0.5} \underset{(>0.999)}{12.14}$ & $\cellcolor[rgb]{0.5,1,0.5} \underset{(<0.001)}{-36.1}$ &  \\ 
   \hline
\end{tabular}
}
\begin{tablenotes}
\footnotesize
\item Diebold Mariano test for the energy score and the first horizon. The main number is the t-statistic value, with the corresponding p-value in brackets. Models M1 to M7 are ARX-OLS, ARX-enet, ARX-IJ, ARX-BiJ, ARX-BiJ-$\mu_{d}$,  ARX-GARCH and  ARX-BiJ-$\mu_{d}$-GARCH, respectively.
\end{tablenotes}
\caption{Diebold-Mariano for Energy score}
\label{tab:ES-DM}
\end{table}

As may be observed in Table \ref{tab:ES-DM}, with ES it is not possible to distinguish between the OLS and enet estimation methods, while with the other three criteria the elastic net estimation method procures significantly better forecasting performance. This happens due to the fact that the performance with the ED is better with the ARX-enet model while the opposite  occurs with the EI. Besides, it is shown that the  ARX-BiJ-$\mu_{d}$ model provides significantly better forecasts, followed by the ARX-BiJ model. As already noted, according to the ES criterion models with bivariate jumps offer significantly better results, which means that our adjustment on the jump diffusion models helps to catch the dependencies efficiently. It is clear that according to this score the assumption of no constant mean of the jump helps with forecasting accuracy. The improvement in incorporating no constant mean to the ARX-BiJ model was not significant when applying the other three criteria. The forecasting performance of the ARX-GARCH model is weaker than expected. This could be because CCC-GARCH structures are more focused on symmetric effects.

\begin{figure}[h!]
\resizebox{0.5\textwidth}{!}{
 \subfloat[Off-peak ARX-enet trajectories]{\includegraphics{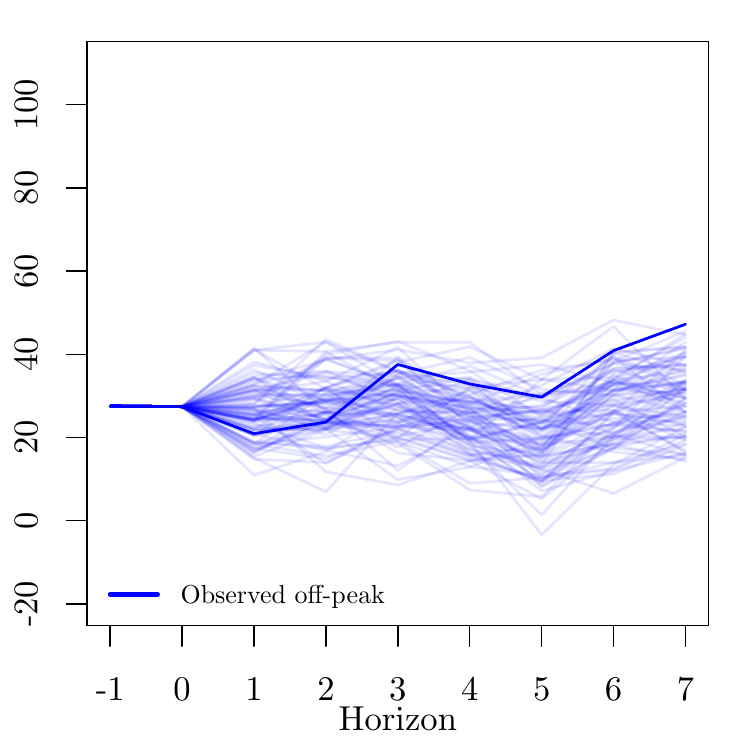}}}
\resizebox{0.5\textwidth}{!}{
 \subfloat[Peak ARX-enet trajectories]{\includegraphics{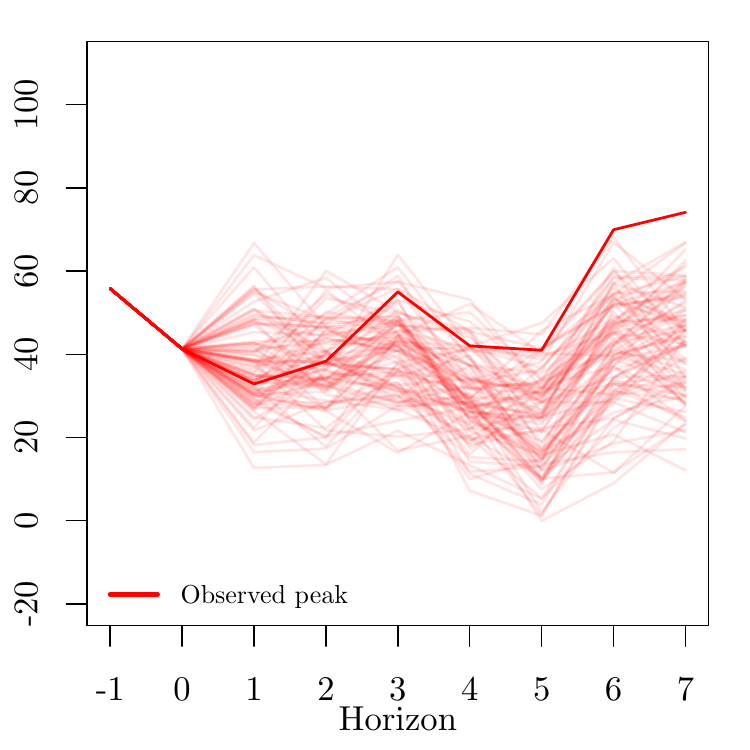}}}\\[-2ex] 
\resizebox{0.5\textwidth}{!}{
 \subfloat[Off-peak ARX-BiJ-$\mu_d$ trajectories]{\includegraphics{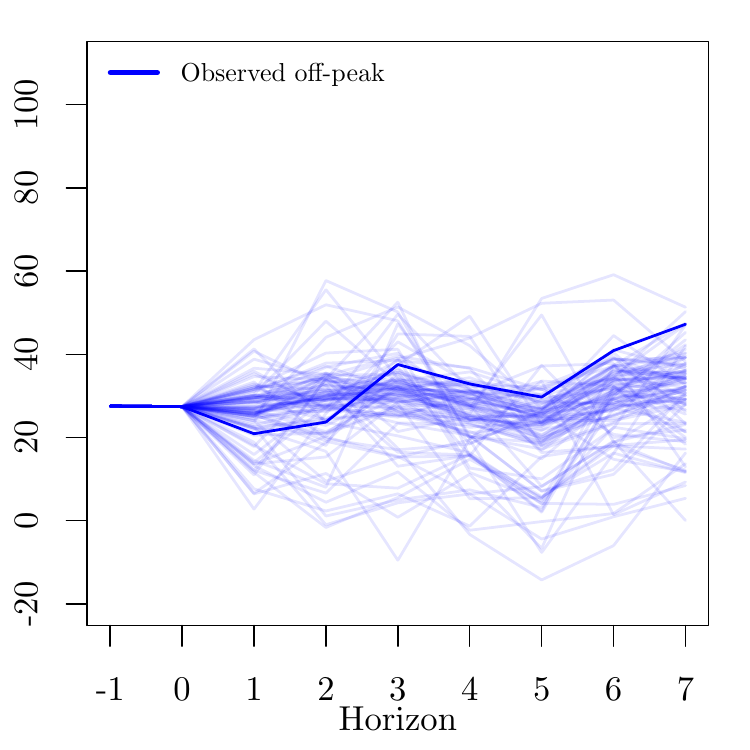}}}
\resizebox{0.5\textwidth}{!}{
 \subfloat[Peak ARX-BiJ-$\mu_d$ trajectories]{\includegraphics{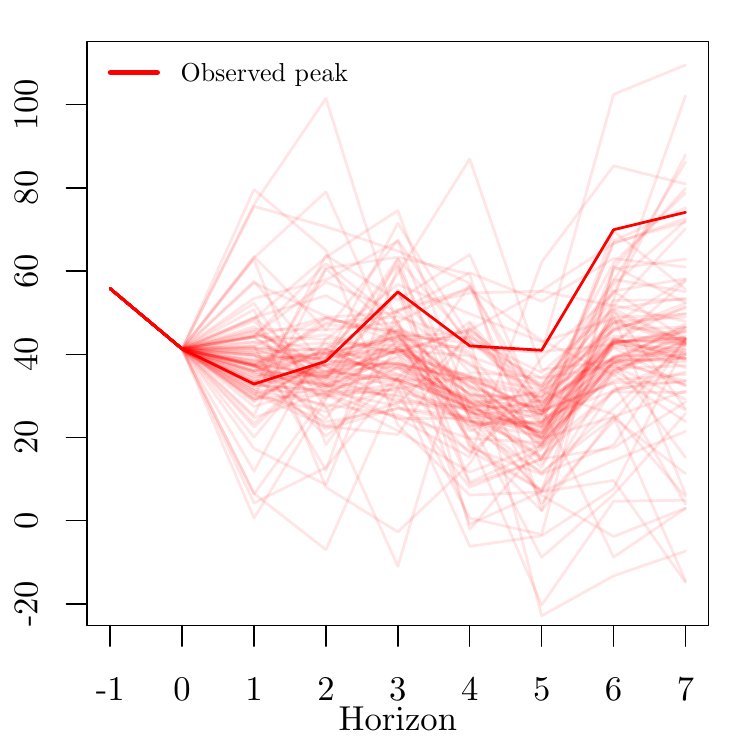}}}
\caption{Trajectories where horizon 0 is 2017-12-12.}
\label{fig:traj2017-12-13}
\end{figure}

\begin{figure}[h!]
 \resizebox{0.5\textwidth}{!}{
 \subfloat[Off-peak ARX-enet trajectories]{\includegraphics{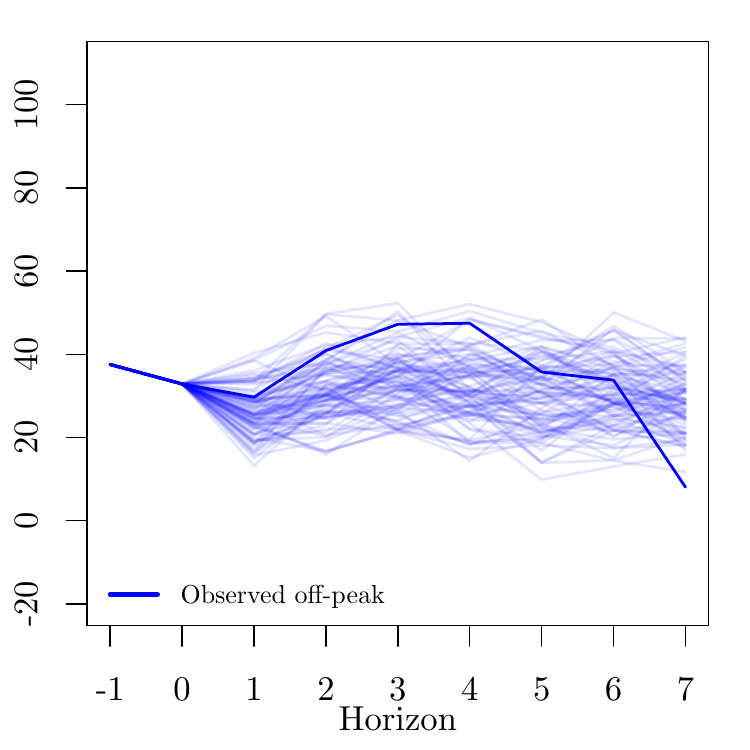}}}
\resizebox{0.5\textwidth}{!}{
 \subfloat[Peak ARX-enet trajectories]{\includegraphics{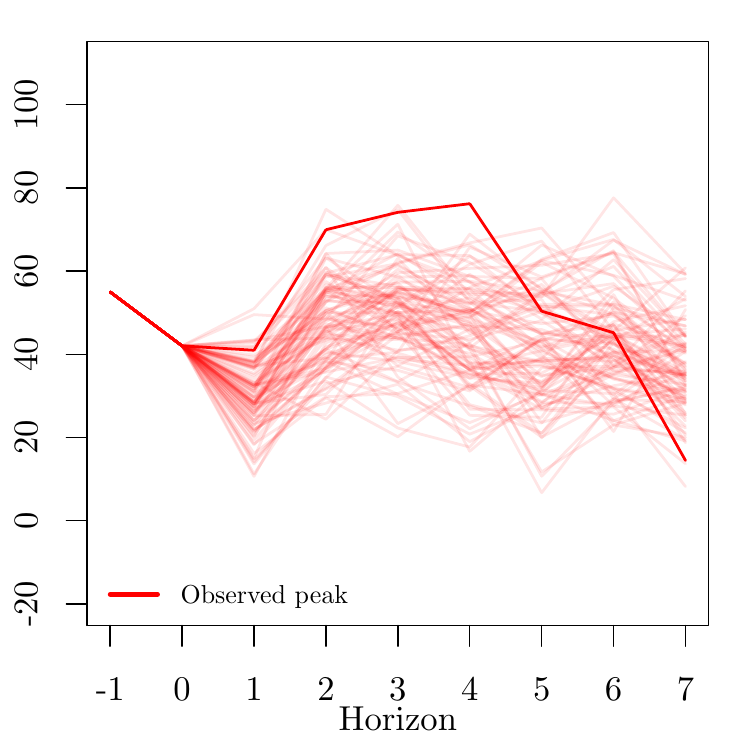}}}\\[-2ex] 
\resizebox{0.5\textwidth}{!}{
 \subfloat[Off-peak ARX-BiJ-$\mu_d$ trajectories]{\includegraphics{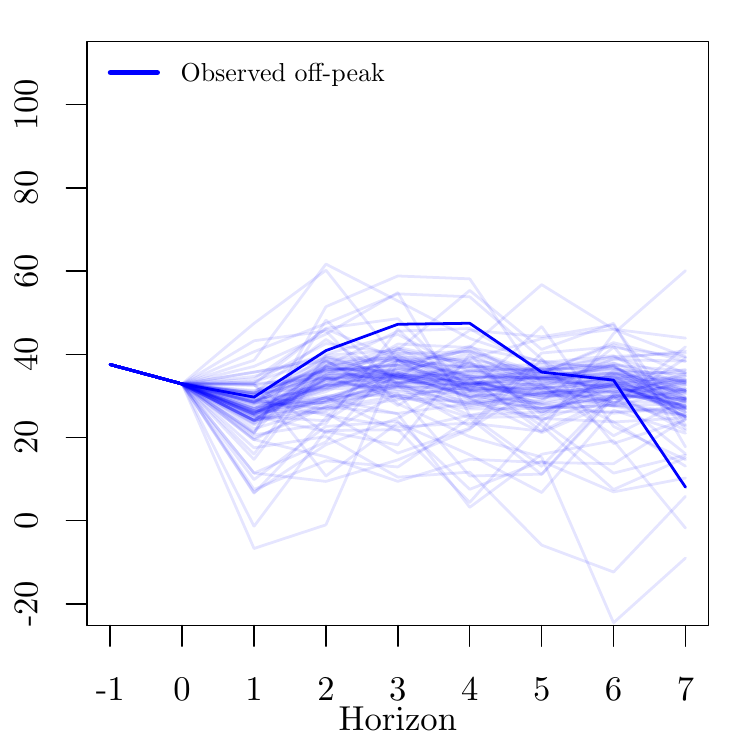}}}
\resizebox{0.5\textwidth}{!}{
 \subfloat[Peak ARX-BiJ-$\mu_d$ trajectories]{\includegraphics{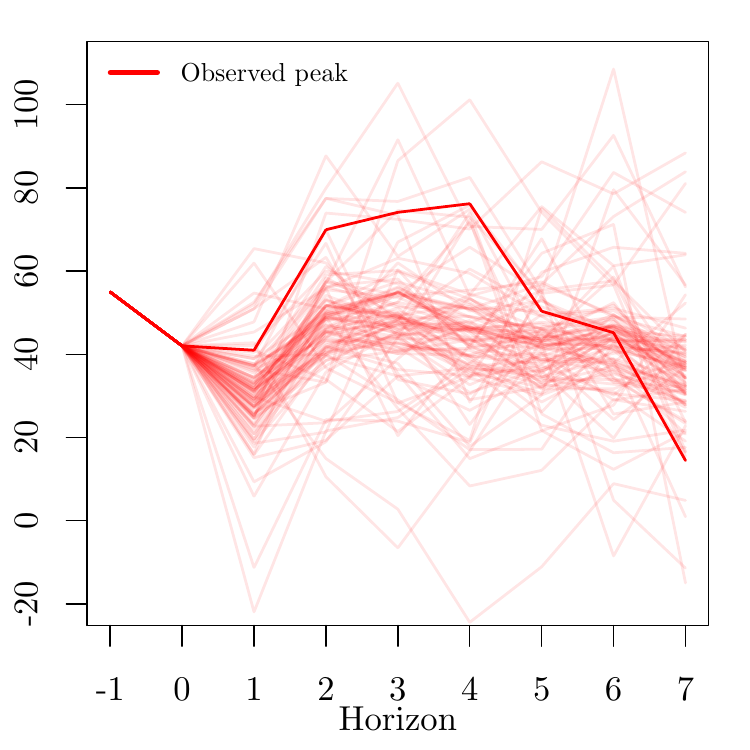}}}
\caption{Trajectories where horizon 0 is 2017-12-16.}
\label{fig:traj2017-12-17}
\end{figure}

In figures \ref{fig:traj2017-12-13} and \ref{fig:traj2017-12-17} we show the first 100 trajectories\footnote{It is not possible to see all the 16000 different trajectories at the same time.} for days 2017-12-12 and 2017-12-16; these days refer to the horizon 0, i.e. when the models are run and the following 7 days are predicted. Only paths for the ARX-enet and ARX-BiJ-$\mu_d$ models are depicted\footnote{It would be tedious for the reader to show paths for all the seven models.}. It may be observed that the trajectories are random and that the observed trajectory or true trajectory is located between the paths most of the times. The shape of the trajectories with the ARX-enet model is more compacted and extreme events are not well captured, some real values are out of the 100 trajectories. On the other hand, it may be observed that the with the ARX-BiJ-$\mu_d$ model the spread of the trajectories is bigger while the trajectories in the middle are more dense. Remember some of the observed values are out of the range of 100 bivariate trajectories of the ARX-enet model, e.g. Figure \ref{fig:traj2017-12-13}. It may be observed in both figures that the correlation between peak and off-peak time series is high as both moves in the same direction. However, in practice it might be highly relevant if price spikes occur at peak and off-peak prices together or not. The ES is the only considered criterion that takes into account the full distribution of the bivariate data. Thus, it is the only criterion which can discriminate for the performance of such a double spike event. It can analyze the whole distribution and it is possible to have a full picture of the predictions. According to the ES, the ARX-BiJ-$\mu_d$ outperforms the ARX-enet model.

The relevance of the ES can be highlighted from the practitioners point of view as well. Imagine there is a trader who manages a portfolio with two assets, peak and off-peak futures contracts. If the portfolio manager is interested in the mean of the portfolio, then the portfolio manager could focus on the MSE criterion. However, the mean behavior is also
included in the ES criterion as full distribution is taken into account. On the other hand, portfolio managers are often interested in the value at risk (VaR) with respect to a certain probability of the portfolio. Then, only the ES is suitable for such an evaluation purpose, as the portfolio return is a weighted sum of two dependent random variables
which depends on the full bivariate distribution. Note that if the peak and off-peak prices would be independent from each other then the resulting distribution could be derived by convolution, and the evaluation by the pinball score on a dense grid would be sufficient. Unfortunately, the prices are highly dependent. Therefore, practitioners should focus on the ES to make decisions regarding the VaR of the portfolio.

We would like to underline that none of the models used in this research is perfect.
The ARX-BiJ-$\mu_d$ model that we claim to be the best one according to the ES criterion
is not perfect as the model can be improved with respect MSE results. If one of the models would be superior to the other ones should be the best model according to all the criteria. 

\section{Conclusion} \label{sec:conclusion}

Proper modeling and forecasting in electricity markets is crucial for all participants. In this paper we focus on off-peak and peak time series. These time series are important for trading in derivatives markets, hence in this paper we try to improve already existing models to achieve accurate forecasts. Participants in derivatives markets can adjust their trading positions and evaluate trading strategies properly when there are accurate forecasts available.

The objective of this paper is to properly incorporate and evaluate the complex dependency structure in bivariate analysis. We believe that it is highly relevant for the forecasts to preserve the correlation structure from the original time series. In the literature, so far, MRJD models have been applied to time series, assuming independence between them. Our approach is to include bivariate jump occurrences in the MRJD model. We then need to assess whether these correlation structures have been properly included or not. To that end, we need a criterion which takes dependencies into account. In our case we use the energy score, but this is not the only criterion we apply: we also use the MAE, MSE, and the pinball score. Additionally, we apply the DM test to compare the models in pairs for each horizon.

It may be observed that models with bivariate jumps do not forecast better according to the MAE and MSE criteria. However, with the pinball score criterion, where the distribution of the forecasts is assessed, the performance is slightly better when a bivariate jump structure is considered in the model. Focusing on the ES score it is shown that models containing correlated jumps perform significantly better compared to those models without them. Nonetheless, the most complex model, which features bivariate jumps, no constant jump size, and CCC-GARCH structure, does not outperform the forecasts of the same model without CCC-GARCH. The way we have chosen the initial values for the ARX-BiJ-$\mu_d$-GARCH model ensures that the in-sample performance is better when CCC-GARCH structures are incorporated. However, the out-of-sample performance is poorer than expected.

For further research it might be interesting to develop dependency structure models considering hourly data and conduct a multivariate analysis with 24 variables. The problem with 24 variables is that the number of parameters to be estimated increases substantially. Here, vine copulas might help to tackle the above mentioned problem.

\section*{Acknowledgements}
Peru Muniain is grateful for financial support from Dpto. de
Educaci\'{o}n, Universidades e Investigaci\'{o}n del Gobierno Vasco under
research grant IT-783-13, from Dpto. de Educaci\'{o}n, Pol\'{\i}tica Ling%
\"{u}\'{\i}stica y Cultura del Gobierno Vasco through Beca Predoctoral de
Formaci\'{o}n de Personal Investigador no Doctor and research grant EGONLABUR from the same department. Peru Muniain  also acknowledges financial support from the Spanish Ministry of Economics and Competitiveness (ECO2015-64467-R MINECO/FEDER).

\bibliography{biblio}

\begin{thebibliography}{}

\bibitem[Bollerslev, 1990]{bollerslev1990modelling}
Bollerslev, T. (1990).
\newblock Modelling the coherence in short-run nominal exchange rates: a
  multivariate generalized {ARCH} model.
\newblock {\em The review of economics and statistics}, pages 498--505.

\bibitem[Cartea and Figueroa, 2005]{cartea2005pricing}
Cartea, A. and Figueroa, M.~G. (2005).
\newblock Pricing in electricity markets: a mean reverting jump diffusion model
  with seasonality.
\newblock {\em Applied Mathematical Finance}, 12(4):313--335.

\bibitem[Dai et~al., 2013]{dai2013multivariate}
Dai, B., Ding, S., Wahba, G., et~al. (2013).
\newblock Multivariate bernoulli distribution.
\newblock {\em Bernoulli}, 19(4):1465--1483.

\bibitem[Diebold, 2015]{diebold2015comparing}
Diebold, F.~X. (2015).
\newblock Comparing predictive accuracy, twenty years later: A personal
  perspective on the use and abuse of {Diebold}--{Mariano} tests.
\newblock {\em Journal of Business \& Economic Statistics}, 33(1):1--1.

\bibitem[Diebold and Mariano, 2002]{diebold2002comparing}
Diebold, F.~X. and Mariano, R.~S. (2002).
\newblock Comparing predictive accuracy.
\newblock {\em Journal of Business \& economic statistics}, 20(1):134--144.

\bibitem[Dudek, 2016]{dudek2016multilayer}
Dudek, G. (2016).
\newblock Multilayer perceptron for {GEFCom2014} probabilistic electricity
  price forecasting.
\newblock {\em International Journal of Forecasting}, 32(3):1057--1060.

\bibitem[Gaillard et~al., 2016]{gaillard2016additive}
Gaillard, P., Goude, Y., and Nedellec, R. (2016).
\newblock Additive models and robust aggregation for {GEFCom2014} probabilistic
  electric load and electricity price forecasting.
\newblock {\em International Journal of forecasting}, 32(3):1038--1050.

\bibitem[Gneiting and Raftery, 2007]{gneiting2007strictly}
Gneiting, T. and Raftery, A.~E. (2007).
\newblock Strictly proper scoring rules, prediction, and estimation.
\newblock {\em Journal of the American Statistical Association},
  102(477):359--378.

\bibitem[Hebiri and Lederer, 2013]{hebiri2013correlations}
Hebiri, M. and Lederer, J. (2013).
\newblock How correlations influence lasso prediction.
\newblock {\em IEEE Transactions on Information Theory}, 59(3):1846--1854.

\bibitem[Higgs, 2009]{higgs2009modelling}
Higgs, H. (2009).
\newblock Modelling price and volatility inter-relationships in the
  {Australian} wholesale spot electricity markets.
\newblock {\em Energy Economics}, 31(5):748--756.

\bibitem[Hoerl and Kennard, 1970]{hoerl1970ridge}
Hoerl, A.~E. and Kennard, R.~W. (1970).
\newblock Ridge regression: Biased estimation for nonorthogonal problems.
\newblock {\em Technometrics}, 12(1):55--67.

\bibitem[Ioannou et~al., 2018]{ioannou2018effect}
Ioannou, A., Angus, A., and Brennan, F. (2018).
\newblock Effect of electricity market price uncertainty modelling on the
  profitability assessment of offshore wind energy through an integrated
  lifecycle techno-economic model.
\newblock In {\em Journal of Physics: Conference Series}, volume 1102, page
  012027. IOP Publishing.

\bibitem[Juban et~al., 2016]{juban2016multiple}
Juban, R., Ohlsson, H., Maasoumy, M., Poirier, L., and Kolter, J.~Z. (2016).
\newblock A multiple quantile regression approach to the wind, solar, and price
  tracks of {GEFCom2014}.
\newblock {\em International Journal of Forecasting}, 32(3):1094--1102.

\bibitem[Karakatsani and Bunn, 2008]{karakatsani2008forecasting}
Karakatsani, N.~V. and Bunn, D.~W. (2008).
\newblock Forecasting electricity prices: The impact of fundamentals and
  time-varying coefficients.
\newblock {\em International Journal of Forecasting}, 24(4):764--785.

\bibitem[Keles et~al., 2012]{keles2012comparison}
Keles, D., Genoese, M., M{\"o}st, D., and Fichtner, W. (2012).
\newblock Comparison of extended mean-reversion and time series models for
  electricity spot price simulation considering negative prices.
\newblock {\em Energy Economics}, 34(4):1012--1032.

\bibitem[Maciejowska and Nowotarski, 2016]{maciejowska2016hybrid}
Maciejowska, K. and Nowotarski, J. (2016).
\newblock A hybrid model for {GEFCom2014} probabilistic electricity price
  forecasting.
\newblock {\em International Journal of Forecasting}, 32(3):1051--1056.

\bibitem[Narajewski and Ziel, 2019]{narajewski2019econometric}
Narajewski, M. and Ziel, F. (2019).
\newblock Econometric modelling and forecasting of intraday electricity prices.
\newblock {\em Journal of Commodity Markets}, page 100107.

\bibitem[Pinson and Girard, 2012]{pinson2012evaluating}
Pinson, P. and Girard, R. (2012).
\newblock Evaluating the quality of scenarios of short-term wind power
  generation.
\newblock {\em Applied Energy}, 96:12--20.

\bibitem[Racine, 2000]{racine2000consistent}
Racine, J. (2000).
\newblock Consistent cross-validatory model-selection for dependent data:
  hv-block cross-validation.
\newblock {\em Journal of econometrics}, 99(1):39--61.

\bibitem[Seifert and Uhrig-Homburg, 2007]{seifert2007modelling}
Seifert, J. and Uhrig-Homburg, M. (2007).
\newblock Modelling jumps in electricity prices: theory and empirical evidence.
\newblock {\em Review of Derivatives Research}, 10(1):59--85.

\bibitem[Silvennoinen and Ter{\"a}svirta, 2009]{silvennoinen2009multivariate}
Silvennoinen, A. and Ter{\"a}svirta, T. (2009).
\newblock Multivariate {GARCH} models.
\newblock In {\em Handbook of financial time series}, pages 201--229. Springer.

\bibitem[Steinert and Ziel, 2019]{steinert2019short}
Steinert, R. and Ziel, F. (2019).
\newblock Short-to mid-term day-ahead electricity price forecasting using
  futures.
\newblock {\em The Energy Journal}, 40(1).

\bibitem[Steinwart et~al., 2011]{steinwart2011estimating}
Steinwart, I., Christmann, A., et~al. (2011).
\newblock Estimating conditional quantiles with the help of the pinball loss.
\newblock {\em Bernoulli}, 17(1):211--225.

\bibitem[Tibshirani, 1996]{tibshirani1996regression}
Tibshirani, R. (1996).
\newblock Regression shrinkage and selection via the lasso.
\newblock {\em Journal of the Royal Statistical Society. Series B
  (Methodological)}, pages 267--288.

\bibitem[Uhlenbeck and Ornstein, 1930]{uhlenbeck1930theory}
Uhlenbeck, G.~E. and Ornstein, L.~S. (1930).
\newblock On the theory of the {Brownian} motion.
\newblock {\em Physical review}, 36(5):823.

\bibitem[Uniejewski et~al., 2016]{uniejewski2016automated}
Uniejewski, B., Nowotarski, J., and Weron, R. (2016).
\newblock Automated variable selection and shrinkage for day-ahead electricity
  price forecasting.
\newblock {\em Energies}, 9(8):621.

\bibitem[Voronin et~al., 2014]{voronin2014hybrid}
Voronin, S., Partanen, J., and Kauranne, T. (2014).
\newblock A hybrid electricity price forecasting model for the {Nordic}
  electricity spot market.
\newblock {\em International Transactions on Electrical Energy Systems},
  24(5):736--760.

\bibitem[Weron, 2008]{weron2008market}
Weron, R. (2008).
\newblock Market price of risk implied by {Asian}--style electricity options
  and futures.
\newblock {\em Energy Economics}, 30(3):1098--1115.

\bibitem[Weron, 2014]{weron2014electricity}
Weron, R. (2014).
\newblock Electricity price forecasting: A review of the state-of-the-art with
  a look into the future.
\newblock {\em International journal of forecasting}, 30(4):1030--1081.

\bibitem[Weron and Ziel, 2019]{weron2019electricity}
Weron, R. and Ziel, F. (2019).
\newblock Electricity price forecasting.
\newblock In U., S. and R., S., editors, {\em Handbook of Energy Economics},
  chapter~35, pages 506--521. Routledge.

\bibitem[Zanotti et~al., 2010]{zanotti2010hedging}
Zanotti, G., Gabbi, G., and Geranio, M. (2010).
\newblock Hedging with futures: Efficacy of {GARCH} correlation models to
  {European} electricity markets.
\newblock {\em Journal of International Financial Markets, Institutions and
  Money}, 20(2):135--148.

\bibitem[Ziel, 2016]{ziel2016forecasting}
Ziel, F. (2016).
\newblock Forecasting electricity spot prices using lasso: On capturing the
  autoregressive intraday structure.
\newblock {\em IEEE Transactions on Power Systems}, 31(6):4977--4987.

\bibitem[Ziel et~al., 2015]{ziel2015efficient}
Ziel, F., Steinert, R., and Husmann, S. (2015).
\newblock Efficient modeling and forecasting of electricity spot prices.
\newblock {\em Energy Economics}, 47:98--111.

\bibitem[Ziel and Weron, 2018]{ziel2018day}
Ziel, F. and Weron, R. (2018).
\newblock Day-ahead electricity price forecasting with high-dimensional
  structures: Univariate vs. multivariate modeling frameworks.
\newblock {\em Energy Economics}, 70:396--420.

\bibitem[Zou and Hastie, 2005]{zou2005regularization}
Zou, H. and Hastie, T. (2005).
\newblock Regularization and variable selection via the elastic net.
\newblock {\em Journal of the Royal Statistical Society: Series B (Statistical
  Methodology)}, 67(2):301--320.

\end{thebibliography}
\bibliographystyle{apalike}

\end{document}